\documentclass[preprint, 12pt]{aastex} 
\usepackage{lscape}

\newcommand{\chandra}{\textit{Chandra}}
\newcommand{\rosat}{\textit{ROSAT}}
\newcommand{\asca}{\textit{ASCA}}
\newcommand{\einstein}{\textit{Einstein}}
\newcommand{\xmm}{\textit{XMM-Newton}}

\newcommand{\fir}{\ensuremath{\mathit f/i}}
\newcommand{\apec}{{\sc apec}}
\newcommand{\xspec}{{\sc xspec}}
\newcommand{\isis}{{\sc isis}}
\newcommand{\spitzer}{\textit{Spitzer}}

\newcommand{\meg}{MEG}

\newcommand{\molecH}{H$_2$}
\newcommand{\Sten}{1--0 S(1)}

\newcommand{\ha}{H$\alpha$}
\newcommand{\lya}{Ly $\alpha$}
\newcommand{\halpha}{\ha}
\newcommand{\kms}{km s$^{-1}$}
\newcommand{\dl}[1]{$\Delta\lambda = #1$}
\newcommand{\trecs}{T-ReCS}
\newcommand{\hd}{HD 283572}

\shorttitle{No disk around DoAr 21?}
\shortauthors{Jensen, Cohen, \& Gagn\'{e}}
\slugcomment{Accepted by {\it The Astrophysical Journal}}

\begin{document}

\title{No Transition Disk? Infrared Excess, PAH, H$_{\rm 2}$, and X-rays
  from the Weak-Lined T Tauri Star DoAr 21}

\author{Eric L. N. Jensen\altaffilmark{1}, David H.
  Cohen\altaffilmark{1}, Marc Gagn\'{e}\altaffilmark{2} }

\altaffiltext{1}{Swarthmore College Department of Physics and
Astronomy, 500 College Ave., Swarthmore, PA 19081. E-mail ejensen1, dcohen1@swarthmore.edu.}
\altaffiltext{2}{West Chester University Department of Geology 
and Astronomy, West Chester, PA 19383}

\begin{abstract}

  As part of a program to understand disk dispersal and the interplay
  between circumstellar disks and X-ray emission, we present new
  high-resolution mid-infrared imaging, high-resolution optical
  spectroscopy, and \chandra\/ grating X-ray spectroscopy of the
  weak-lined T Tauri star DoAr 21.  DoAr 21 (age $< 10^6$ yr and mass
  $\sim 2.2$ M$_\sun$ based on evolutionary tracks) is a strong X-ray
  emitter, with conflicting evidence in the literature about its disk
  properties.  It shows weak but broad \ha\ emission (reported here
  for the first time since the 1950s); polarimetric variability; PAH
  and \molecH\ emission; and a strong, spatially-resolved 24-\micron\
  excess in archival {\it Spitzer\/} photometry.  Gemini
  sub-arcsecond-resolution 9--18 \micron\ images show that there is
  little or no excess mid-infrared emission within 100 AU of the star;
  the excess emission is extended over several arcseconds and is quite
  asymmetric.  The extended emission is bright in the UV-excited
  $\lambda= 11.3$ \micron\ PAH emission feature.  A new
  high-resolution X-ray grating spectrum from \chandra\/ shows that
  the stellar X-ray emission is very hard and dominated by continuum
  emission; it is well-fit by a multi-temperature thermal model,
  typical of hard coronal sources, and shows no evidence of unusually
  high densities.  A flare during the X-ray observation shows a
  temperature approaching $10^8$ K\null.  We argue that the
  far-ultraviolet emission from the transition region is sufficient to
  excite the observed extended PAH and continuum emission, and that
  the \molecH\ emission may be similarly extended and excited.  While
  this extended emission may be a disk in the final stages of
  clearing, it also could be more akin to a small-scale
  photodissociation region than a protoplanetary disk, highlighting
  both the very young ages ($< 10^6$ yr) at which some stars are found
  without disks, and the extreme radiation environment around even
  late-type pre--main-sequence stars.
  
\end{abstract}


\section{Introduction}
\label{section:introduction}

An open question in early stellar evolution and planet formation is
the process and timescale by which stars disperse their protoplanetary
disks.  Central to this question is the lifetime of the gas component
of circumstellar disks.  The gas dominates the disk mass, but since
small dust grains dominate the infrared opacity, dispersal or growth
of these grains can significantly reduce or eliminate the excess
infrared emission that is often taken to be the primary signature of a
circumstellar disk, even if substantial gas is still present.

Related to the question of gas disk survival and evolution is the
influence of X-ray emission on the circumstellar gas, and vice versa.
T Tauri stars, low-mass pre--main-sequence stars, are often strong
X-ray emitters \citep[see, e.g., the review of] [and references
therein]{FeigelsonMontmerle1999}.  A detailed understanding of the
X-ray emission from these young stars is important both for an
understanding of the early evolution of the stars themselves, and due
to the X-rays' impact on the surrounding environment.  T Tauri stars
are often surrounded by disks of gas and dust, from which planets
presumably form later in the pre--main-sequence phase.  The physical
state of these disks is strongly influenced by the star's X-ray
emission.  The X-rays are the dominant source of ionization of the
disk \citep{IgeaGlassgold1999}, they strongly influence the disk
chemistry \citep{Maloney1996}, and they can be important in disk
photoevaporation \citep{GortiHollenbach2009,Drake2009}.  X-ray flares
from the young Sun may have produced the isotopic anomalies seen today
in meteorites \citep{Feigelson2002}.

Similarly, the presence of circumstellar gas may play a role in
shaping or modifying the star's X-ray emission.  The model for the
X-ray emission of the vast majority of T Tauri stars is that the
X-rays are produced by solar-type magnetic activity scaled up by rapid
rotation and/or longer convective turnover times
\citep[e.g.,][]{FeigelsonMontmerle1999,Preibisch2005}.  In a small
number of systems, there is also evidence that the X-rays are
influenced by interaction between a stellar magnetic field and a
circumstellar disk, perhaps via accretion \citep{Kastner2002,
  StelzerSchmitt2004, Schmitt2005}.  The unprecedented X-ray spectral
resolution and sensitivity of the \chandra\/ X-ray satellite offers an
opportunity to place much better constraints on the temperature and
density of the X-ray emitting gas, thereby providing potentially
important information about the X-ray emission mechanism(s), and about
the X-ray radiation field incident on the circumstellar disk.

To date, however, only a modest number of high-spectral-resolution
X-ray observations of pre--main-sequence stars have been published;
even the distances to the nearest star-forming regions result in X-ray
fluxes that make high-resolution observations prohibitively long.
There are a handful of notable exceptions, however.  TW Hya
\citep{Kastner2002} and HD 98800 \citep{Kastner2004} both lie in the
nearby TW Hya association, with a distance of roughly 50 pc and an age
of 5--15 Myr \citep[][and references therein]{Weintraub2000}.  The
\chandra\/ X-ray spectrum of TW Hya yielded a surprisingly low
temperature for the X-ray emitting gas, and suggestions of very high
densities in the X-ray emitting regions, leading \citet{Kastner2002}
to suggest that the X-ray emission is related to accretion from the
star's circumstellar disk.  \citet{StelzerSchmitt2004} reached a
similar conclusion based on density diagnostics in TW Hya's \xmm\/
spectrum, and also attributed the low Fe abundance seen in the X-ray
emitting gas to depletion of some elements from the gas phase onto
dust grains in the disk.  \citet{Schmitt2005} presented
similar evidence for high-density X-ray emitting gas around the
classical T Tauri star BP Tau.

In contrast, the X-ray spectra of both HD 98800 \citep{Kastner2004}
and the similarly-aged ($\sim 10$ Myr) PZ Tel \citep{Argiroffi2004}
are quite similar to spectra of older solar-type stars and show no
evidence for unusually high densities, suggesting that their X-ray
activity is, to first order, scaled-up solar activity.  Solar-like
X-ray activity is also seen in AB Dor \citep{Sanz-Forcada2003}, which
is near or recently arrived on the zero-age main sequence.  Notably,
TW Hya and BP Tau show clear evidence of active accretion from their
circumstellar disks, while the stars with more solar-like X-ray
emission do not have disks.\footnote{Only the southern component (HD
  98800 B) of the wide binary pair in HD 98800 has circumstellar
  material \citep{Koerner2000}, while \citet{Kastner2004} show clearly
  that the X-ray spectrum they detect and analyze is from the diskless
  northern component, HD 98800 A.}  In general, while accretion may
modify the X-ray spectrum in some young stars, studies of the X-ray
emission from large samples of young stars show that accretion is not
a primary driver of X-ray activity \citep{Preibisch2005,Stassun2006}.

In the context of understanding both disk evolution and the interplay
between disks and X-rays, the weak-lined T Tauri star DoAr 21 (a.k.a.\
V2246 Oph, Elias 2-14, VSSG 23, GSS 23, ROXs 8, Haro 1-6, YLW 26, HBC
637, ROXR1 13) presents an interesting case study.  It has no strong
evidence for accretion or a massive disk, and yet it is one of the few
T Tauri stars around which emission from \molecH\ and polycyclic
aromatic hydrocarbon (PAH) molecules has been detected (Sec.\
\ref{section:circumstellar}).  \citet{Bary2002} suggest that it may
retain a gas-only disk in which all of the dust has coagulated into
larger grains; \citet{Cieza2008} classify it as a ``transition disk,''
one in the process of being cleared from the inside out.  In addition,
it is the most luminous X-ray source in $\rho$ Oph, making it an
excellent laboratory to study the influence of X-rays and
circumstellar material.

\begin{figure}
\begin{center}
\includegraphics[scale=0.75]{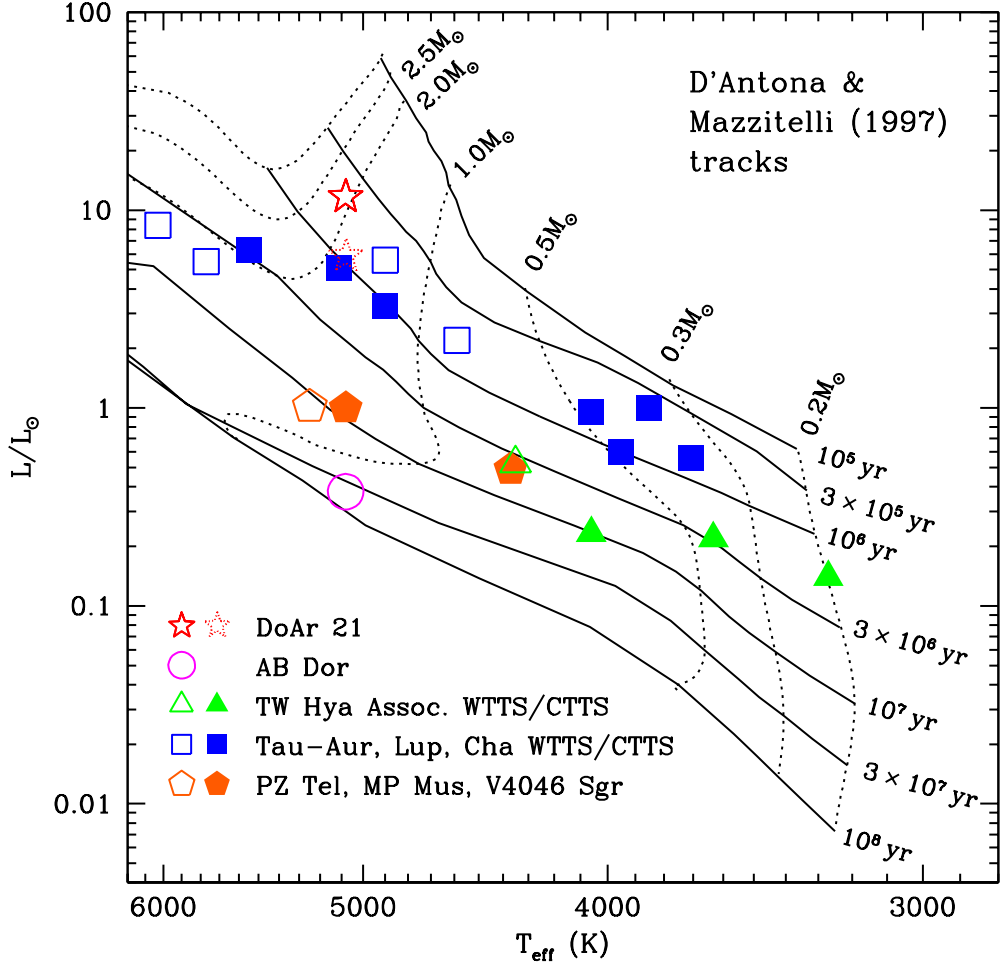}
\end{center}
\caption{HR diagram showing the position of DoAr 21 (star symbols)
  compared to other pre--main-sequence stars that have been observed
  with the \chandra\/ or \xmm\/ gratings.  The dotted star shows the
  luminosity of each star in DoAr 21 if it is an equal-mass binary.
  Filled symbols are classical T Tauri stars, and open symbols are
  weak-lined T Tauri stars.  DoAr 21 is both younger ($\sim 4 \times
  10^5$ yr) and more massive ($\sim 2.2$ M$_\sun$) than most of the T
  Tauri stars observed previously.  The pre--main-sequence
  evolutionary tracks shown are from \citet{dm97}.  Both the
  \citet{ps99} and \citet{Siess2000} tracks result in slightly older
  absolute ages in general, but similar relative ages.}
\label{figure:hr}
\end{figure}

Here we present new Gemini mid-infrared images, high-resolution
optical spectra, and \chandra\/ high-resolution grating spectra of
DoAr 21. The primary contrast of DoAr 21 with the handful of
pre--main-sequence stars previously observed at high X-ray spectral
resolution (Fig.\ \ref{figure:hr}) is the combination of its youth
and lack of a strong accretion signature.  DoAr 21 is embedded in the
$\rho$ Oph star-forming region (see \citealt{Wilking2008} for a recent
review), whereas a number of the others (e.g., TW Hya, Hen 3-600, MP
Mus, and V4046 Sgr, with ages of $\sim 10$--30 Myr) no longer lie in
regions of active star formation, with all surrounding molecular gas
having been dissipated.  With an estimated age of $\lesssim 1$ Myr,
DoAr 21 is younger than these other stars.  A few of the stars
observed with high X-ray spectral resolution (e.g.  T Tau, BP Tau, RU
Lup) are of a similar age to DoAr 21, but all are classical T Tauri
stars with strong accretion signatures. The central questions here,
then, are how much difference the young age and lack of strong
accretion make in the X-ray emission of DoAr 21; what the distribution
and physical conditions of its circumstellar material are; and how the
high-energy radiation field incident on the circumstellar material may
influence its properties.

Below we discuss the stellar and circumstellar properties of DoAr 21
(Sec.\ \ref{section:stellar}), showing new images of its circumstellar
material, which, surprisingly, is not present near the star but which
extends to several hundred AU\null. We present new high-resolution
optical spectra which show that DoAr 21 is probably not accreting, but
which are consistent with strong chromospheric activity.  In this
light, we then present our new X-ray observations (Sec.\
\ref{section:x-ray}) and analyze the X-ray emission in detail through
emission-line-ratio fits, modeling of the entire spectrum, and
analysis of time variability of the emission before, during, and after
a large flare.  In Sec.\ \ref{section:discussion}, we discuss the
influence of the star's X-ray and ultraviolet emission on its
circumstellar environment.  Based on the level of activity suggested
by our X-ray observations, we estimate the far-ultraviolet (FUV) flux,
suggesting that it is sufficient to excite the observed PAH emission
even at hundreds of AU from the star.  Finally, we present a possible
geometry for the \molecH\ emission, suggesting that it may be more
akin to a photo-dissociation region (PDR) than a circumstellar disk.

\section{DoAr 21's stellar properties and circumstellar environment}
\label{section:stellar}

Before considering our new X-ray data for DoAr 21, here we consider
new and existing data from other wavelengths, including new
mid-infrared imaging and high-resolution optical spectroscopy.  We
also consider infrared and optical photometry and spectroscopy from
the literature in order to build up a more complete picture of the
stellar properties and circumstellar environment of DoAr 21.

\subsection{Stellar properties and multiplicity}
\label{section:mult}

Spectral types in the literature are K0 \citep{Bouvier1992,
  Martin1998} from optical spectra and K0--K2 \citep{Luhman1999} from
near-infrared spectra.  Following \citet{Luhman1999}, we adopt a
spectral type of K1 and effective temperature of 5080
K\null.\footnote{The spectral type of B2 found in some references, and
  as of this writing in Simbad, is from \citet{Chini1977}, and is
  based on B, R, I, and K photometric colors.  The authors did not
  find a unique combination of spectral type and extinction that fit
  the data; assuming an extinction of $A_V = 10.1$ they found a
  spectral type of B2, and assuming $A_V = 6.6$--6.2 they found a
  spectral type of G5--K0.  A follow-up paper by \citet{Chini1981}
  used a broader range of colors and found $A_V = 6.2$ and a spectral
  type of K0, consistent with what we find here.}  The distance to
DoAr 21 from a VLBA parallax measurement is $122 \pm 6$ pc
\citep{Loinard2008}.  Using this distance and our fit to the
de-reddened spectral energy distribution (Sec.\ \ref{section:sed}), we
estimate the stellar luminosity to be $11.7 \pm 0.8$ $L_\sun$, where
the uncertainty arises primarily from the large photometric
variability and somewhat uncertain extinction.

The VLBA observations of \citet{Loinard2008} show DoAr 21 to be a
binary, with a projected separation of 5 mas (0.6 AU) in one of their
observations.  The orbital parameters are uncertain, but Loinard et
al.\ estimate the semimajor axis to be 1--2 AU\null.  Since the binary
pair has been detected only at radio wavelengths thus far, there is no
information about the mass ratio or the optical luminosity ratio.
Ascribing all of the luminosity to the primary star (solid-lined star in
Fig.\ \ref{figure:hr}) yields a mass of $\sim 2.2$ $M_{\sun}$ and an
age of $\sim 4\times 10^5$ yr. If the luminosity is split equally
between the two stars (dotted star in Fig.\ \ref{figure:hr}), the mass
of each star is $\sim 1.8$ $M_{\sun}$ and the age is $\sim 8\times
10^5$ yr.  There is some evidence in the optical spectra (Sec.\
\ref{section:opticalspectrum}) that the secondary is contributing to
the observed line width, suggesting that its optical luminosity is
non-negligible; thus the latter set of parameters may be closer to
reality.  We note that DoAr 21 is relatively massive among T Tauri
stars, and it has a K spectral type only because of its very young
age.  By an age of 3--4 Myr, it will move into the spectral type range
of Herbig Ae/Be stars, and it will arrive on the main sequence as a
late B or early A star.

\subsection{Spectral energy distribution and infrared excess}
\label{section:circumstellar}
\label{section:optical}\label{section:infrared}\label{section:sed}

The evidence in the literature for circumstellar material around DoAr
21 (and accretion of such material) is mixed, and DoAr 21 has often
been considered
\citep[e.g.,][]{LadaWilking1984,Andre1992,Preibisch1999,Bontemps2001}
to be a Class III or diskless T Tauri star, though some sources have
stated that it has an infrared excess \citep[e.g.,][]{Bouvier1992},
and it has recently been classified as a ``transition disk'' system,
with a cleared inner disk but optically thick outer disk
\citep[e.g.,][]{Cieza2008}.

In light of the lack of a strong infrared excess, the detection of
emission features from PAHs \citep*{Hanner1995} and \molecH\
\citep{Bary2003,Bitner2008} was surprising, raising the possibility
that a substantial gas component of the disk is still present.  Early
observations showed H$\alpha$ with bright and variable emission
\citep{Haro1949}, and then later weakly in emission
\citep{Dolidze1959, Hidajat1961}.
\label{section:h-alpha-history} Since then, however, no published
observations of the star have shown H$\alpha$ emission
\citep{Rydgren1976, Montmerle1983, Feigelson1985, WSB1987,
  Bouvier1992, Martin1998}. Br$\gamma$ is in absorption with no
near-infrared veiling \citep{Luhman1999}.  Our new high-resolution
optical spectra (Sec.\ \ref{section:opticalspectrum}) show that the
photospheric H$\alpha$ absorption line is partly filled in with
emission and is surrounded by weak but very broad ($\sim 300$ \kms\
FWHM) emission wings, revealing a weak and variable emission component
that is consistent with the \ha\ emission seen in older,
coronally-active low-mass stars and which thus suggests that there is
little to no accretion onto DoAr 21.

\begin{figure}
\begin{center}
\includegraphics[angle=-90,scale=0.5]{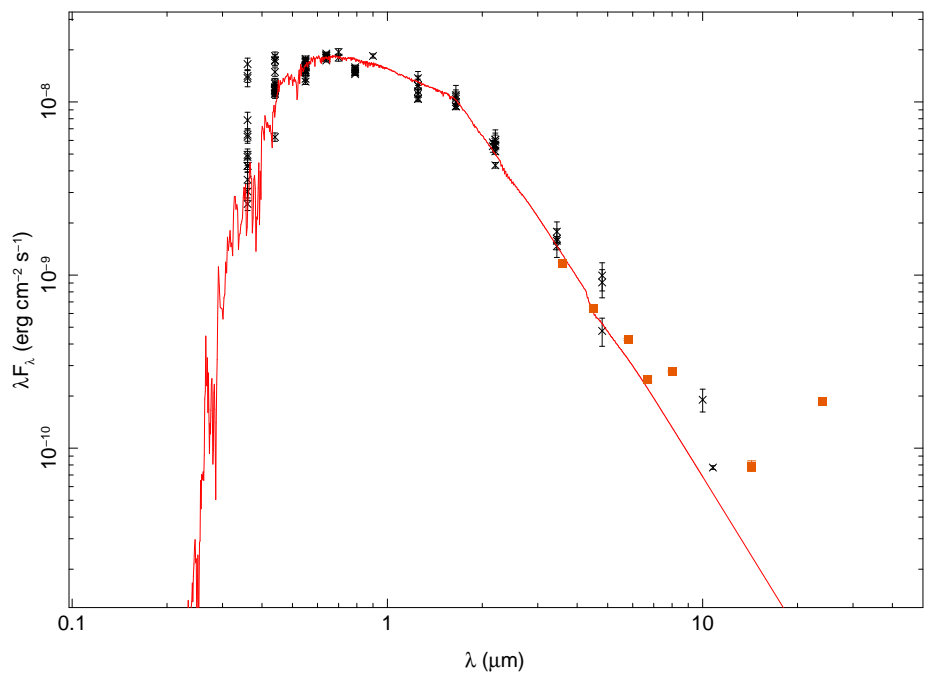}
\end{center}
\caption{The spectral energy distribution of DoAr 21, from ultraviolet
  through infrared wavelengths. Spitzer and ISO data are shown as
  filled squares (orange in the on-line edition), and the 850 \micron\
  and 1.3 mm upper limits are not shown.  The photometry has been
  de-reddened with $A_V = 6.2$ and $R_V = 4.2$ as described in the
  text.  Overplotted is a model photosphere (solid line); the star
  shows a clear infrared excess at $\lambda \ge 8$ \micron.}
\label{figure:sed}
\end{figure}

The spectral energy distribution (SED) of DoAr 21 gives some clues to
the nature of its circumstellar environment.  We searched the
literature for available photometry of DoAr 21, and assembled the SED
from $\lambda = 0.39$ \micron\ ($U$ band) to $\lambda = 1.3$ mm.  The
observed broad-band colors are much redder than photospheric colors
for a K1 star.  To de-redden the photometry and to determine whether
or not DoAr 21 shows any evidence of infrared excess, a signature of
the presence of a circumstellar or circumbinary disk, we compared the
SED of DoAr 21 to a model photosphere \citep{BuserKurucz1992} with
$T_{\rm eff} = 5000$ K (Fig.\ \ref{figure:sed}).  We adopt the
extinction law of \citet{wd01}, with the modifications of
\citet{Draine2003}.  This extinction law has been shown to be a better
fit to mid-infrared data than a power-law extrapolation from the
near-infrared \citep{RomanZuniga2007, Chapman2008, McClure2009}.  We
use a ratio of total-to-selective extinction $R_V \equiv {A_V}/
{E(B-V)} = 4.2$, determined from the relationship $R_V = 5.7
\lambda_{\rm max}$ \citep{Vrba1993}, where $\lambda_{\rm max} = 0.74$
\micron\ is the wavelength of maximum observed polarization for DoAr
21 \citep{Martin1992}. This value of $R_V$ is consistent with results
found for other lines of sight in Ophiuchus \citep{Vrba1993} and
yields a notably better fit to the $UBVRI$ data than the standard
interstellar value of $R_V = 3.1$. Physically, a larger $R_V$
corresponds to larger grain sizes, as expected in a dense cloud.
Using this extinction law, we de-reddened the photometry to find the
best fit of the model photosphere to the $V$, $R$, and $I$ data, as
the $U$ and $B$ data show significant variability and the
longer-wavelength data may be affected by any infrared excess.  This
procedure yielded $A_V = 6.2$ mag, and combined with our adopted
distance, a photospheric luminosity of $11.7 \pm 0.8$ $L_\sun$.  This
corresponds to a stellar radius of $4.6 \pm 0.2$ $R_\sun$ if all of
the luminosity is attributed to one star.  The adopted $A_V$
corresponds to $A_J = 1.8$.  We note that this is consistent with the
neutral hydrogen column density of $N_H \approx 10^{22}$ atoms
cm$^{-2}$ found from the X-ray spectrum (Sec.\ \ref{section:x-ray});
from a sample of 20 sources with good X-ray and infrared data in the
$\rho$ Oph cloud, \citet{Vuong2003} find $N_H / A_J = 5.6 \pm 0.4
\times 10^{21}$ cm$^{-2}$ mag$^{-1}$.  Given the differences in
extinction laws used, our adopted extinction is reasonably consistent
with that found by other authors; \citet{Luhman1999} find $A_J = 1.6$
mag, and \citet{Bouvier1992} find $A_V = 6.6$.  The de-reddened
photometric data are plotted in Figure \ref{figure:sed}; the
millimeter-wavelength upper limits of \citet{Andre1990} and
\citet{Cieza2008} are not shown.


While the observed optical and near-infrared variability (see Section
\ref{section:opt-variability}) makes it impossible to fit every data
point, the photometry is broadly consistent with the model photosphere
within the uncertainties for $\lambda \le 5$ \micron.  There is some
evidence for excess emission at $U$ and $B$, which may be due to
flaring as discussed below. While the brightest $U$ and $B$ points are
clearly inconsistent with photospheric emission, the fainter points
lie close to the model photosphere, consistent with a scenario in
which DoAr 21 is flaring much of the time.

At longer wavelengths, there is a clear infrared excess at $\lambda =
7$--24 \micron.  These points are quite insensitive to extinction
corrections, since the extinction is so small, and to the
determination of the star's effective temperature, since they lie on
the Rayleigh-Jeans tail of the photospheric emission.  The space-based
infrared data are highlighted in the figure as squares (orange in the
on-line edition), as they are not affected by the Earth's atmosphere
and are measured on a uniform photometric system; the $\lambda = 6.7$
and 14 \micron\ points are ISO data from \citet{Bontemps2001}, while
the other $\lambda = 3.6$--24 \micron\ squares are \spitzer\ IRAC and
MIPS data from \citet{Cieza2007}.  While the presence of an excess
shortward of 7 \micron\ is debatable, the 8--24 \micron\ points are
well in excess of the photospheric emission, indicating the presence
of some circumstellar material within a radius of
4\arcsec--9\arcsec, the aperture size of the mid-infrared
observations.  This circumstellar material apparently does not have a
large mass in small dust grains, however, given the very sensitive
$3\sigma$ flux upper limits of $< 4$ mJy at $\lambda = 1300$ \micron\
\citep{Andre1990} and $< 18$ mJy at $\lambda=850$ \micron\
\citep{Cieza2008}.

\subsection{Mid-infrared imaging}

\subsubsection{Archival Spitzer  images}
\begin{figure}
\begin{center}
\includegraphics[scale=1.3]{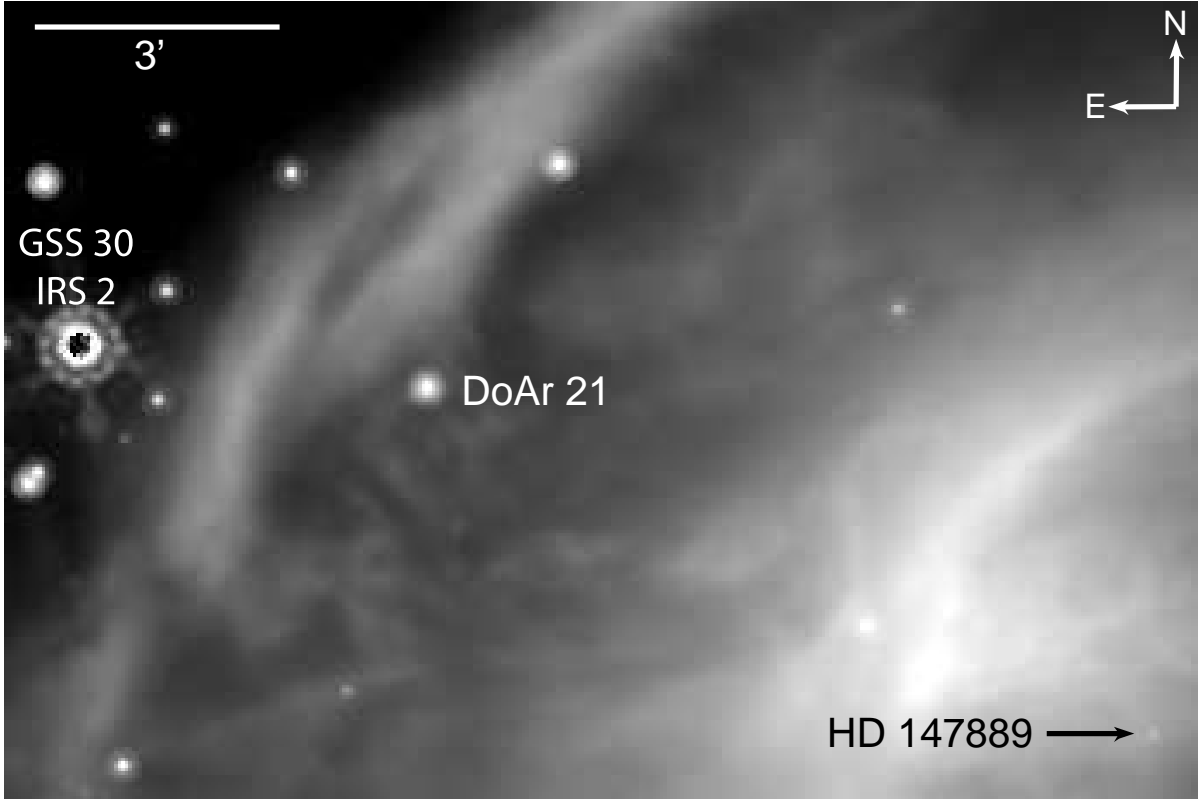}
\end{center}
\caption{{\it Spitzer\/} MIPS 24-\micron\ image of the field around DoAr
21; DoAr 21 is clearly detected, indicating a substantial amount of
circumstellar material. Unlike other stellar sources in the field,
DoAr 21 does not show the first Airy ring, indicating that the
24-\micron\ emission is marginally resolved.  The B2 star HD 147889
(lower right) excites the emission from the bright ridge to the northeast of DoAr 21.
\label{figure:spitzer-image}}
\end{figure}

DoAr 21 was observed by the {\it Spitzer\/} Space Telescope with both
IRAC ($\lambda = 3.6$, 4.5, 5.8, 8.0 \micron) and MIPS ($\lambda =
24$, 70, and 160 \micron) as part of the c2d Legacy program
\citep{Cieza2007,Padgett2008}.  PSF-fitting photometry of DoAr 21 from
3.6--24 \micron\ is reported by \citet{Cieza2007} and plotted in
Figure \ref{figure:sed}; here we concentrate on the images themselves.
Figure \ref{figure:spitzer-image} shows the $\lambda = 24$ \micron\
image of the region around DoAr 21; DoAr 21 itself is clearly
detected.  Visual examination of the $\lambda=70$ \micron\ images
shows a source at the position of DoAr 21, but the noise is too high
for accurate photometry.

Surprisingly, the 24-\micron\ emission from DoAr 21 is somewhat
extended.  Close inspection of the 24-\micron\ MIPS images reveals
that the profile of DoAr 21 is different from that of other stars in
the field.  Since \spitzer\ is diffraction-limited at 24 \micron, the
point-spread-function (PSF) is a modified Airy pattern, and indeed
many of the sources in Figure \ref{figure:spitzer-image} (including
sources that are fainter than DoAr 21 and/or projected on comparable
or brighter background emission) show the presence of the first Airy
ring.  DoAr 21 does not, however, indicating that it is not a point
source at 24 \micron.  To check this, we compared the
azimuthally-averaged radial profile of DoAr 21 to that of a star in
the same image (GSS 26), and also to a sample empirical on-orbit PSF
(constructed from many stars in a MIPS mosaic image of the Trapezium)
supplied by the Spitzer Science Center (SSC).  DoAr 21 is clearly more
extended than the other stars, both of which show the signature of the
first Airy ring at radii of 3--5 pixels, while DoAr 21 does not.  (The
diffraction limit of \spitzer\ at 24 \micron\ is 7\farcs4, while the
MIPS mosaic pixel size is 2\farcs45, so the first null occurs at a
radius of 3 pixels.)  We examined three different MIPS mosaic images,
and DoAr 21 appears extended in all of them.  We also examined the
individual baseline-calibrated data (BCD) frames from the SSC archive
(from which the mosaics were constructed), and found that DoAr 21
appears extended in each frame, regardless of its position on the
chip, while other stars in the frame do not.  Thus, the broader
profile of DoAr 21 is not an artifact of the mosaicing process.

To estimate the radial extent of the emission, we convolved an
oversampled MIPS PSF with Gaussians of different radii, resampled them
at the MIPS pixel scale, and compared their radial profiles to that of
DoAr 21.  The profile of DoAr 21 is roughly consistent with a Gaussian
source with a sigma of 0.75--1 MIPS pixel (1\farcs8--2\farcs5).  At
the 122 pc distance of DoAr 21, this corresponds to a FWHM of 500--700
AU\null.  Thus, the 24-\micron\ emission appears to be extended to
radii of 250--350 AU from the star.

\subsubsection{New Gemini images}
\label{section:gemini-data}

Motivated by the extended emission seen in the \spitzer\ images, we
observed DoAr 21 with the \trecs\ camera on Gemini South in June and
July 2007.  The observations were made through filters with central
wavelengths (and 50\% transmission widths) in \micron\ of 8.6
(\dl{0.43}), 10.4 (\dl{1.02}), 11.3 (\dl{0.61}), and 18.3 (\dl{1.51}).
The 8.6 and 11.3 \micron\ filters are centered on PAH features, while
the 10.4 \micron\ Si-4 and 18.3 \micron\ Qa filters largely measure
continuum emission.  In particular, the Si-4 filter acts as a nearby
continuum reference for any PAH emission.

The data were taken with the standard chop and nod technique to remove
sky background emission, and were processed with the standard Gemini
IRAF pipeline.\footnote{IRAF is distributed by the National Optical
  Astronomy Observatories, which are operated by the Association of
  Universities for Research in Astronomy, Inc., under cooperative
  agreement with the National Science Foundation.}  The chop throw was
15\arcsec\ at a position angle of 130\arcdeg, parallel to the ridge of
24-\micron\ emission seen in the \spitzer\ images (Fig.\
\ref{figure:spitzer-image}); this chop direction was chosen to
minimize variation of the background emission.

The data were flux-calibrated with observations of HD 136422 (K5 III)
at a similar airmass and near in time to the DoAr 21 observations in
each filter.  Fluxes for HD 136422 were taken from the calibrated
template spectrum created by \citet{Cohen1999}.  Because the target
and calibrator were observed at similar, low ($\lesssim 1.3$)
airmasses and near in time to each other, we did not attempt to
correct for differential atmospheric extinction between target and
calibrator observations.

The spatial resolution of our observations, as measured by the FWHM of
standard stars observed at similar airmass to DoAr 21, was
0\farcs35--0\farcs38 in the 8.6--11.3 \micron\ observations, and
0\farcs58 at 18.3 \micron. This is near the diffraction limit of
Gemini South, and is well-sampled by the \trecs\ pixel scale of
0\farcs09 pixel$^{-1}$.

\begin{figure}
\begin{center}
\includegraphics[scale=0.73]{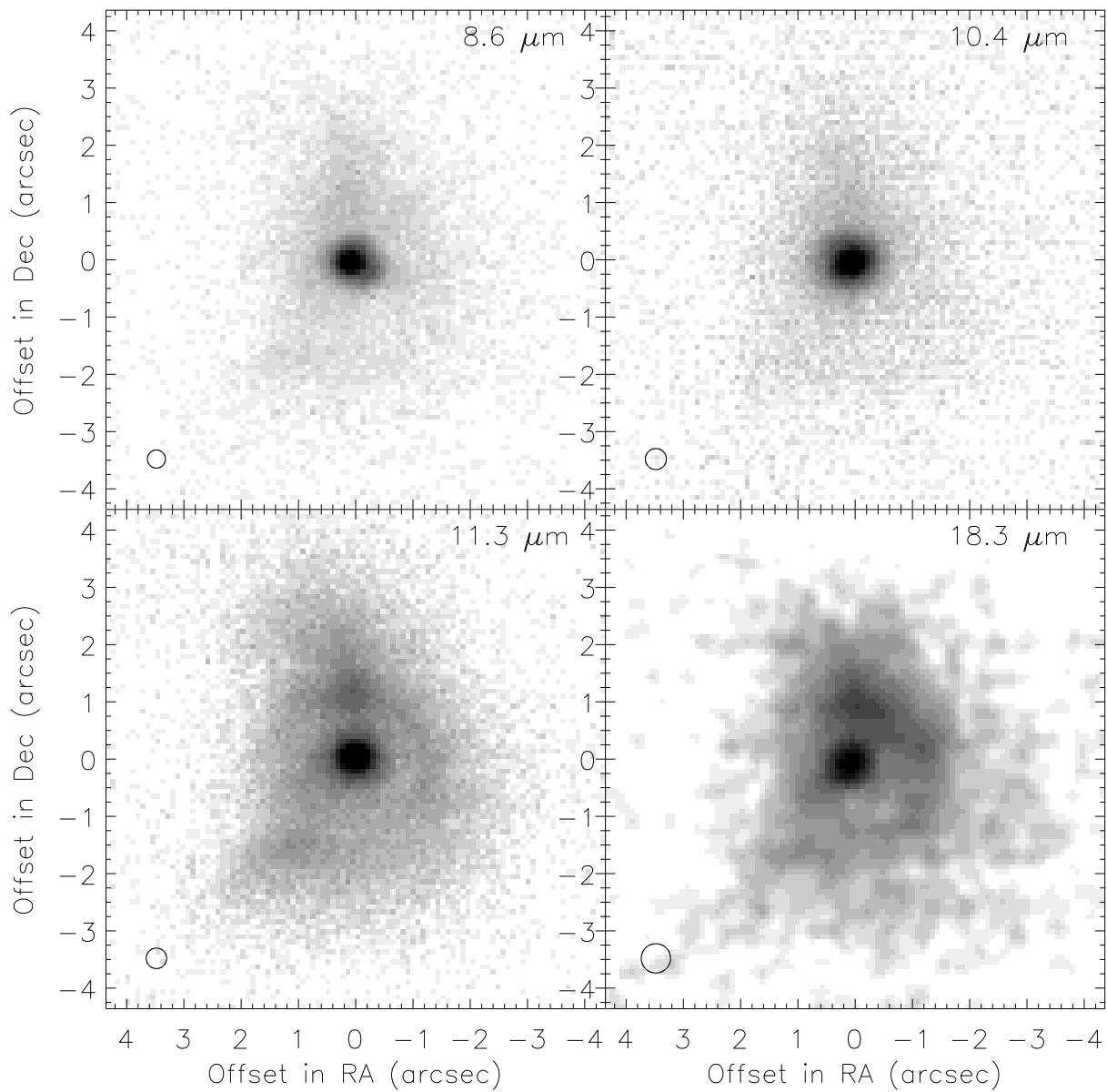}
\end{center}
\caption{Gemini T-ReCS images of DoAr 21 at four different
  wavelengths.  The 8.6- and 11.3-\micron\ bands are
  centered on PAH emission features, while the other two bands trace
  continuum emission.  The grayscale shows the emission displayed with 
  a logarithmic stretch; the 18.3-\micron\ image is smoothed 
  with a 3-pixel FWHM Gaussian.  The circle in the lower left of each
  image shows the FWHM of the PSF measured from a comparison star
  observed at similar airmass. 
  \label{figure:gemini-image}}
\end{figure}

The resulting images are shown in Figure \ref{figure:gemini-image}.
Irregular, extended emission is seen at all wavelengths.  The emission
is brighter in the filters containing PAH features (8.6 and especially
11.3 \micron) than in the 10.4 \micron\ continuum, suggesting that the
extended material shows PAH emission.  The brightest off-source
emission is a spot roughly 1\farcs1 north of DoAr 21, and there is a
ridge or partial arc of emission to the north and west, seen in all
filters.

The central stellar source in the 8.6--11.3 \micron\ DoAr 21 images
appears point-like at our resolution; it has a FWHM similar to that of
the photometric calibrator in the corresponding band. The slight
extension of the PSF at PA $\sim 230$\arcdeg\ in the 8.6 \micron\
images is also seen in the calibrator. The central source in the 18.3
\micron\ image has FWHM of 0\farcs6--1\farcs0; given the much lower
signal-to-noise ratio at this wavelength, it is difficult to tell if
the central source is extended, though certainly there is significant
extended emission beyond this.

\begin{figure}
\begin{center}
\includegraphics[scale=0.8]{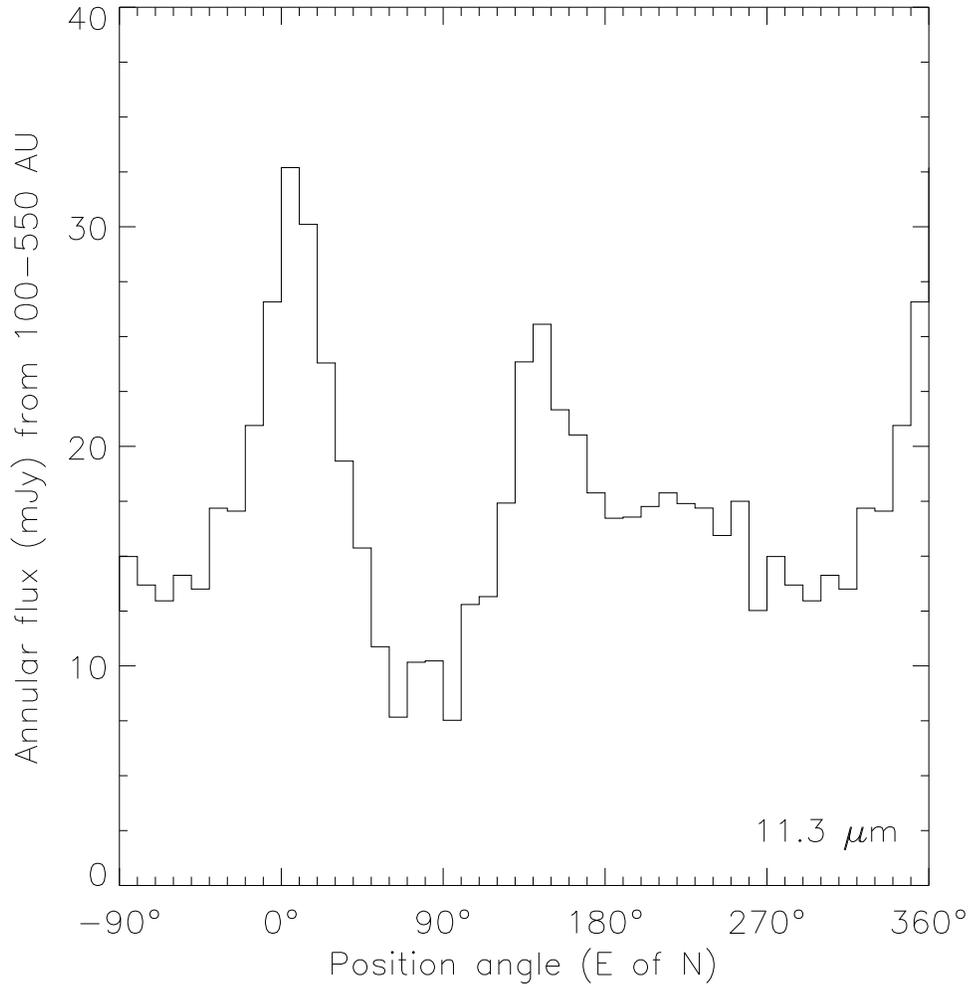}
\end{center}
\caption{Azimuthal profile of the 11.3-\micron\ emission between radii
  of 100 AU and 550 AU, showing the 4:1 contrast in brightness at
  different position angles.  The profiles in the 8.6
  \micron\ and 10.4 \micron\ images show similar structure and
  contrast, though at lower flux levels.
  \label{figure:gemini-azimuthal-profile}}
\end{figure}

The extended emission is quite asymmetric, unlike the emission
expected from a disk.  There is a partial arc or ring to the north and
west (seen most prominently at 11.3 and 18.3 \micron) at roughly
1\farcs1 from the star.  There are bright knots at position angles of
roughly 0\arcdeg\ and 140\arcdeg, with the northern knot appearing
both brighter and more concentrated, and the SE knot being visible to
larger radii, especially at 18.3 \micron.  As shown in the
star-subtracted azimuthal profile in Figure
\ref{figure:gemini-azimuthal-profile}, there is a roughly 4:1 contrast
between the brightest and faintest position angles.

Archival HST WFPC/2 images of DoAr 21 in the F606W and F814W filters
show a faint filament extending 8\arcsec--10\arcsec\ southeast from the star;
the filament is not apparent in images in the F1042M filter.  There is
no obvious brightness gradient along the filament, so
\citet{Stapelfeldt2009} suggest that it is foreground or background
material illuminated by the star, rather than a jet associated with
the star.  This filament is also visible in the narrow-band
near-infrared images of \citet{Gomez2003}, in filters both on and off
the \molecH\ \Sten\ line.  Comparison of the WFPC/2, near-IR, and
Gemini images shows that this filament overlaps the extension seen in
Figure \ref{figure:gemini-image} at PA 140\arcdeg.  Thus, the
circumstellar material apparently extends to even larger radii than
seen in our images, at least in some directions.  We discuss this
point further in Section \ref{section:discussion}.

\begin{figure}
\begin{center}
\includegraphics[angle=-90,scale=0.65]{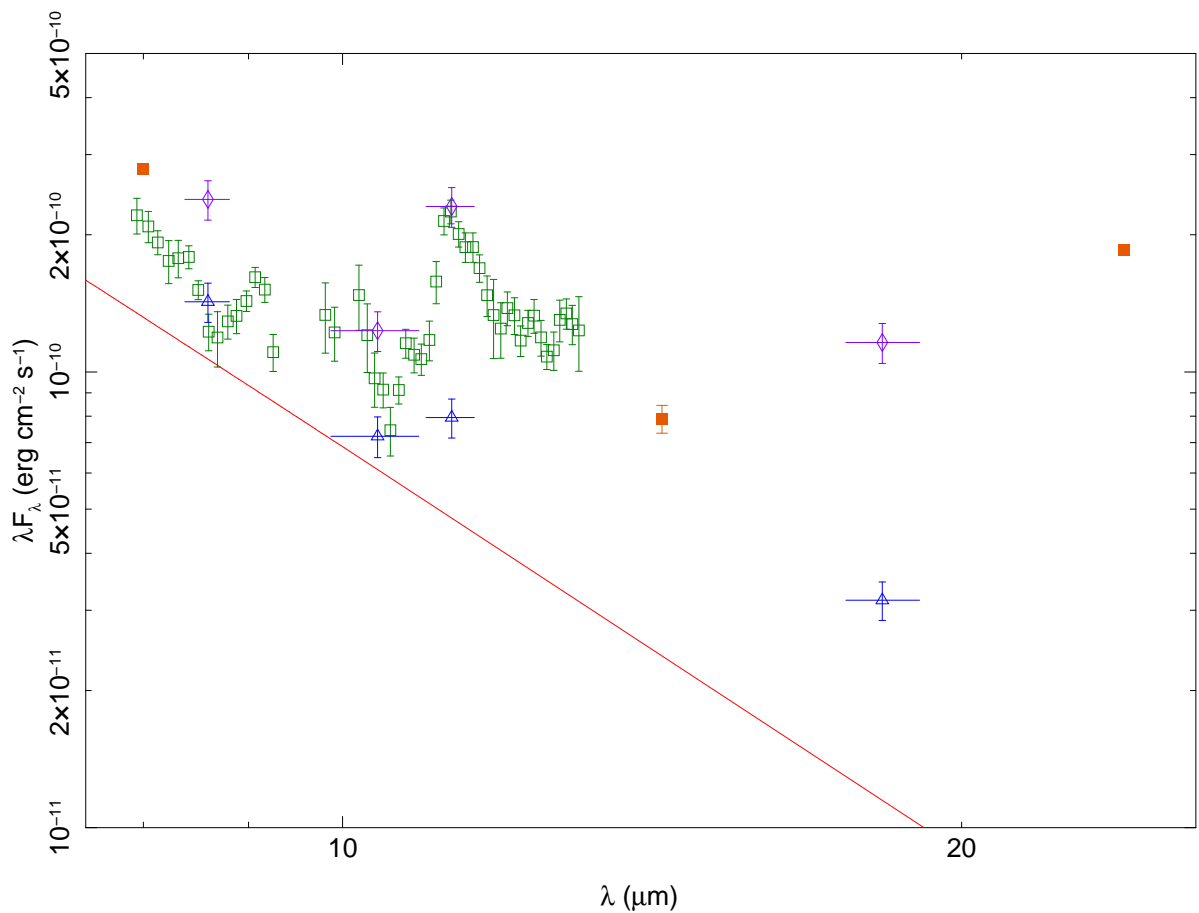}
\caption{A close-up of the mid-infrared SED, with the model
  photosphere and de-reddened photometry as in Figure
  \ref{figure:sed}.  The broad-band Spitzer and ISO data are shown as
  filled squares (orange in the on-line edition), and the spectrum
  from \citet{Hanner1995} is shown as open squares (green in the
  on-line edition), showing an emission feature at 11.3 \micron\
  attributed to PAHs.  De-reddened photometric measurements from new
  Gemini narrow-band images (wavelengths 8.6, 10.4, and 11.3 \micron)
  are overplotted, with horizontal bars showing the bandwidth of the
  filters used and vertical error bars showing the assumed 10\%
  uncertainty.  The open triangles (blue in the on-line edition) show
  measurements within 100 AU (0\farcs82) of the star, showing little
  or no excess in the continuum at 10.4 \micron.  The open diamonds
  (purple in the on-line edition) show measurements in a 2\farcs75
  radius aperture, the same as that used by \citet{Hanner1995}.}
\label{figure:pah}
\end{center}
\end{figure}

To quantify the spatial distribution of circumstellar material, and in
particular to see whether there is any infrared excess emission that
is not spatially resolved, we measured the flux in our images in
apertures of different radii.  We chose aperture radii of 0\farcs82,
2\farcs75, and 6\arcsec.  The smallest radius corresponds to a
physical radius of 100 AU, a typical disk radius for a young star; the
intermediate radius matches the aperture used by \citet{Hanner1995}
for their mid-infrared spectroscopy; and the largest radius
encompasses all the detected flux above the background level in our
images, probing roughly the same spatial scale as the MIPS 24-\micron\
images\null.  These fluxes are quoted in Table \ref{table:gemini-photometry}
and plotted in Figure \ref{figure:pah}.  There is little infrared
excess within 100 AU of the star, while the photometry of the more
extended emission is roughly consistent with previous photometry.

\begin{center}
\begin{deluxetable}{lccc}
\tablewidth{0pt}
\tablecaption{Photometry of DoAr 21\label{table:gemini-photometry}
}
\tablehead{
  \colhead{Filter} & \colhead{Flux within  0\farcs82} &
 \colhead{Flux within 2\farcs75} &  \colhead{Flux within 6\arcsec}\\
\colhead{} & \colhead{(mJy)} &\colhead{(mJy)}&\colhead{(mJy)}}
\startdata
\phn 8.6 \micron\ (PAH)  & 313  & 525  & 596 \\
10.4 \micron\  (Si-4)      & 176  & 300  & 347 \\
11.3 \micron\ (PAH) & 234  & 680  & 816 \\
18.3 \micron\ (Qa)       & 165  & 607  & 728 \\
\enddata
\end{deluxetable}
\end{center}

Given the clear presence of a significant amount of circumstellar
material, but evidence that little of it is close to the star, we now
turn to the question of whether there is any evidence for accretion
onto DoAr 21.

\subsection{Optical spectrum}
\label{section:rotation}\label{section:opticalspectrum}
\label{section:opticalobs}

As noted in Section \ref{section:h-alpha-history}, the \ha\ emission
from DoAr 21 has shown varying behavior at different times in the
past, and thus additional spectroscopic observations are useful to
probe this behavior.  In addition, DoAr 21's stellar rotation period
is unknown, as it did not show periodic photometric variability in the
observations of \citet{Bouvier1988}; thus, measurement of its
projected rotational velocity $v \sin i$ is useful.

To probe the \ha\ emission and stellar rotation of DoAr 21, we
observed it with the echelle spectrograph on the CTIO Blanco 4-meter
telescope on 2002 June 12 and 14, and on the four consecutive nights
2003 April 11--14.  The spectrograph setup gave wavelength coverage of
approximately 5,000--8,000 \AA\ with 0.08 \AA/pixel.  With a 1\arcsec\
slit width, this gave a spectral resolving power (as measured from the FWHM
of narrow lines in a ThAr comparison lamp spectrum) of $R = 40,000$ at
the H$\alpha$ line.  Exposure times were 20 minutes on 2002 June 12,
15 minutes on 2003 April 11 and 30 minutes on the other four nights.
The data were reduced using standard routines for echelle spectra in
IRAF.

Our spectra show broad absorption lines.  There is no obvious evidence
in the spectra of doubled lines from a binary companion, though the
broad lines and relatively low signal-to-noise ratio limit our ability
to detect other lines.  The Li I 6708 \AA\ absorption line has an
equivalent width of $320 \pm 20$ m\AA, consistent with the measurement
of \citet{Martin1998}.  From our spectra we measure a
radial velocity of $-6 \pm 4$ km s$^{-1}$, consistent with the
velocity of $-4.6 \pm 3.3$ km s$^{-1}$ measured by
\citet{Massarotti2005}, and with the $\rho$ Oph young-star mean of
$-6.3 \pm 1.5$ km s$^{-1}$ \citep{Prato2007}.  Comparison of our
spectra with a K1 V standard spectrum artificially broadened to
various velocities gives a projected rotational velocity $v \sin i =
80 \pm 10$ km s$^{-1}$ for DoAr 21. Combined with the stellar radius
of 4.6 R$_\sun$ estimated from its effective temperature and
luminosity (attributing all the luminosity to one star), this gives a
rotational period of around 3 days, or less if the star is not viewed
equator-on.\footnote{\citet{Bary2003} suggest that the inclination
  DoAr 21 is greater than 55\arcdeg.  If this is the case, then the
  equatorial rotational velocity is greater than 100 km s$^{-1}$.}

Given the discovery that DoAr 21 is a binary (Sec.
\ref{section:mult}), however, there is an alternate interpretation
of the width of its spectral lines.  \citet{Massarotti2005} report $v
\sin i = 29$ km s$^{-1}$ for DoAr 21.  The origin of the discrepancy
between their determination and ours is unclear, but it could result
from observing a binary at different orbital phases, so that sometimes
the lines are broadened (though not clearly doubled) due to different
radial velocities of the two stars.  The \citet{Massarotti2005}
spectra, measured at similar spectral resolution but with less
spectral coverage and lower S/N, show some variation in the width of
the cross-correlation peak, which is a measure of the line width,
occasionally showing broader peaks (D. Latham, personal
communication).  Our spectra show consistently broad lines, and none
of our six spectra is consistent with $v \sin i$ as low as 29 km
s$^{-1}$.  We repeated the procedure used by \citet{Massarotti2005} on
our spectra, cross-correlating various artificially broadened template
spectra with DoAr 21 to find the template that gives the highest
correlation peak, and we consistently find $v \sin i \approx$ 80 km
s$^{-1}$.  However, our spectra cover only two windows of a few days
each, so it is possible that both our 2002 and 2003 observations were
obtained at orbital phases with relatively large velocity differences.
If we assume an equal-mass binary, the evolutionary tracks in Figure
\ref{figure:hr} give masses of $\sim 1.8$ $M_{\sun}$ for each star, or
a total system mass of $\sim 3.6$ $M_{\sun}$.  With an assumed
semimajor axis of 1--2 AU \citep{Loinard2008}, the relative orbital
velocity of the two stars in a circular orbit would be $\sim $40--60
km s$^{-1}$, which is of the right magnitude to account for the
difference in line broadening observed in our spectra and those of
\citet{Massarotti2005}.  The inferred orbital period of 0.5--1.5 yr
from these orbital parameters is also consistent with our two
observations (separated by ten months) falling either half or a whole
orbital period apart, so that the same velocity difference would be
observed.

\begin{figure}
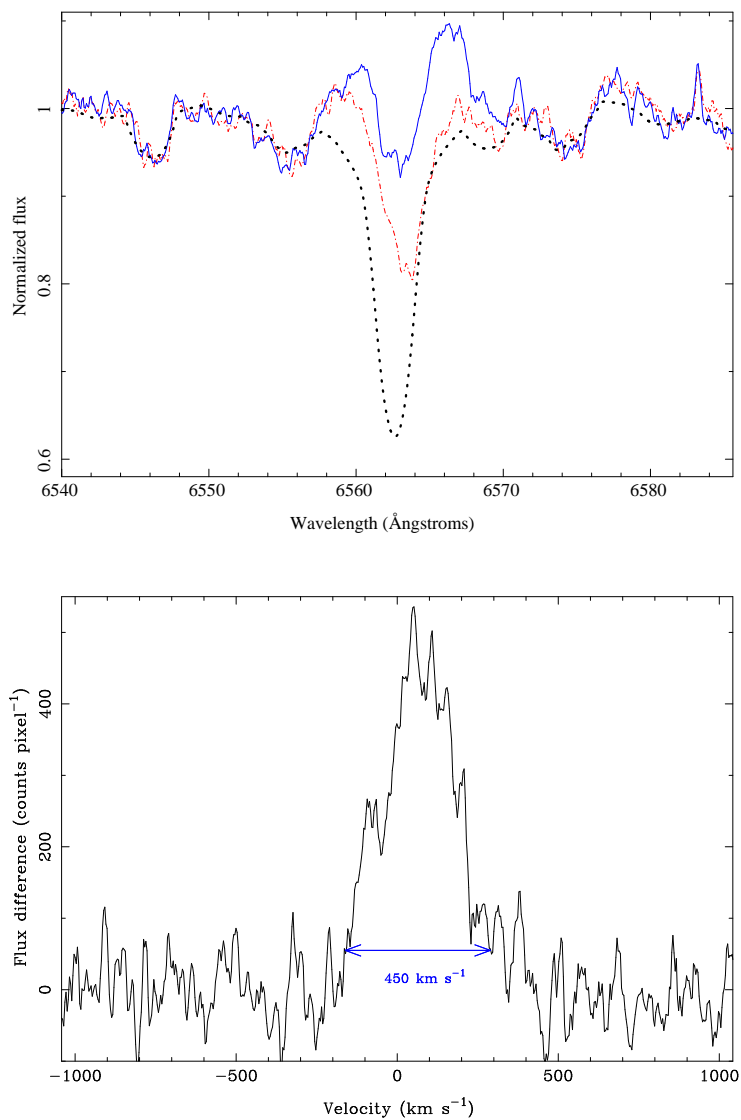

\begin{center}
\includegraphics[angle=-90,scale=0.4]{figure7a.ps}\\[8mm]
\includegraphics[angle=-90,scale=0.4]{figure7b.ps}
\end{center}
\caption{High-resolution optical spectra of DoAr 21 in the vicinity of
  the \ha\ line. Top: The dot-dashed line (red in the online edition)
  shows DoAr 21 on 12 April 2003; the solid line (blue in the online
  edition) shows DoAr 21 on 13 April 2003; the dotted line shows the
  K1 V standard star HD 13445, artificially broadened to $v \sin i =
  80$ km s$^{-1}$.  The \ha\ line in DoAr 21 clearly is filled in by a
  broad emission component that is present on both nights, but
  stronger on the second night.  Bottom: The difference between the
  two DoAr 21 spectra, showing the broad emission component that
  appears the second night.  The full-width at 10\% intensity is
  indicated.}
\label{figure:halpha}
\end{figure}

DoAr 21 has at times shown \ha\ in emission in the past
\citep{Haro1949,Dolidze1959,Hidajat1961}, while other observations
have shown \ha\ absorption.  Our observations, shown in Figure
\ref{figure:halpha}, show a combination of both.  There is a central
photospheric absorption line, but our high spectral resolution reveals
the presence of broad emission wings.  In addition, comparison with
the K1 V spectral standard HD 13445 \citep{KeenanMcNeil1989},
artificially broadened to $v \sin i = 80$ km s$^{-1}$, shows that DoAr
21's \ha\ absorption line is substantially filled in with emission.
After correcting for the underlying photospheric absorption, the
emission component we observe in DoAr 21 has an emission equivalent
width of 1--2 \AA.

Further, the line varies substantially from one night to the next.
Figure \ref{figure:halpha}b shows the difference between two nights,
clearly showing the presence of a broad, variable emission component
at \ha.  The emission shown here is a lower limit to the true \ha\
emission line flux, since any emission component present on both
nights is subtracted out.  The FWHM of the emission is $\sim 300$
\kms, while the full width at 10\% intensity is $\sim 450$ \kms.
\citet{WhiteBasri2003} propose that stars with \ha\ full width at 10\%
intensity of $> 270$ \kms\ should be considered classical T Tauri
stars, arguing that the line velocity width is a better accretion
diagnostic than the more commonly used equivalent width; they observe
a range of line widths from 270 to 670 \kms\ in accreting sources.  By
this measure, DoAr 21 could be considered to be a classical
(accreting) T Tauri star, and indeed \citet{Bary2002} suggest that
DoAr 21 is still accreting material from a disk based on their
detection of \molecH.

While it is possible that DoAr 21 could have some low-level accretion,
its \ha\ emission is also consistent with a lack of accretion, since
some stars that are not thought to be accreting show \ha\ behavior
similar to that seen in DoAr 21.  \citet{Montes1997} find that some
chromospherically active binary stars show \ha\ emission that is
well-fit by a broad component and a narrow component, with the broad
component having FWHM values from 133 to 470 \kms.
\citet{Fernandez2004} find that the quiescent (non-flare) spectrum of
the weak-lined pre--main-sequence triple system V410 Tau has \ha\ that
is just filled in, not overtly in emission, while during a flare the
\ha\ equivalent width is 27 \AA.  After subtraction of a photosphere
of the correct spectral type from the quiescent spectrum, they find
that the residual \ha\ emission line (equivalent width $\sim 1$ \AA,
similar to the emission component from DoAr 21) has a broad component
with a FWHM of 300--400 \kms.  During a flare, the H$\beta$ line is
observed to be in emission with FWHM of 730 \kms, with its width
declining as the flare decays.  V410 Tau shows no evidence of infrared
excess or other indicators of accretion, though definitive detection
of a very weak infrared excess might be difficult given the shortage
of resolved infrared measurements of the three stars (all within
0\farcs3; \citealt{WhiteGhez2001}) in the system.  There is no
detectable emission in the CO fundamental \citep{Najita2003} nor
\molecH\ FUV lines \citep{Herczeg2006}.  With no evidence of either a
gas or a dust disk around V410 Tau, the \ha\ seems more likely to be
related to the flares than to accretion.  Thus, broad Balmer emission
lines are not by themselves an unambiguous indicator of accretion.

We revisit the question of what the \halpha\ and \molecH\ emission
tell us about DoAr 21's circumstellar material in Section
\ref{section:discussion}.   Here we simply note that the chromospheric
activity implied by one interpretation of the \halpha\ emission, and
the need for a strong UV flux to excite the \molecH\ and PAH emission,
both imply a need for a better understanding of the high-energy
emission from DoAr 21.  Thus, we now turn to an examination of our new
high-resolution X-ray spectrum of DoAr 21.

\section{X-ray emission}
\label{section:x-ray}

DoAr 21 is the brightest X-ray source in the $\rho$ Oph cloud core A,
with an X-ray luminosity of nearly $10^{32}$ ergs s$^{-1}$, a hard
thermal spectrum, and large X-ray flares seen in several archival
observations at a rate of nearly one per day. Its X-ray emission has
been observed at low resolution with \einstein\/
\citep{Montmerle1983}, \rosat\/ \citep{Casanova1995}, \asca\/
\citep{Koyama1994}, and \chandra\/ \citep{itk2002,gsd2004}.

The new \chandra\/ grating spectra we present here have vastly
superior spectral resolution compared to any of these previous
datasets, enabling us to extract temperature and abundance information
with more reliability and also, for the first time, to examine density
diagnostics.  Given the high rate of flaring seen in previous
observations, it is not surprising that we detected a large flare in
our 94 ksec \chandra\/ observation.  Thus, we can bring the same
temperature and abundance diagnostic tools to bear on the pre-flare,
flare, and post-flare portions of the new dataset.

The High Energy Transmission Grating Spectrometer (HETGS) has two
grating arrays: the Medium Energy Grating (MEG), with a FWHM
resolution of 2.3 m\AA; and the High Energy Grating (HEG), with a
resolution of 1.2 m\AA\/ \citep{Canizares2005}. Both grating arrays
operate together, with the dispersed spectra (first, second, and third
orders) as well as the zeroth order spectrum recorded on the ACIS CCD
array. We used standard CIAO (v.3.3) tools to extract the dispersed
spectra, and to create observation-specific spectral response matrices
and effective area tabulations, as well as to create light curves and
also to extract spectra for each of the three subsets of the
observation. Of the dispersed spectra, only the first order spectra
have a significant number of counts.  The zeroth order spectrum is
severely piled up, so we restrict the analysis we report on here to
the first order MEG and HEG spectra.  We performed most of the
spectral analysis using \xspec\ v.12.3.

The goal of our X-ray spectral and timing analysis is to characterize
the properties of the hot plasma on this very magnetically active
pre--main-sequence star, and to compare its properties to those
measured in other pre--main-sequence stars with high-resolution X-ray
spectra.  Its extreme youth and lack of obvious signatures of
accretion make for an interesting contrast between DoAr 21 and the
strongly accreting T Tauri stars that have been well-studied with the
\chandra\/ gratings \citep{gt2007}. Although DoAr 21 is not a
classical T Tauri star, it does---as we have detailed in the previous
section---have some circumstellar material in its vicinity.  The
observed X-rays, and the unobservable far- and extreme-UV emission
associated with the X-ray emitting plasma as it cools, could have an
important effect on this circumstellar material, perhaps being the
dominant source of excitation for the PAH and \molecH\ emission.
Thus, characterizing the X-ray properties of the star is important for
understanding the physical conditions in the circumstellar
environment.

\begin{figure*}
\begin{center}
\includegraphics[angle=0,width=160mm]{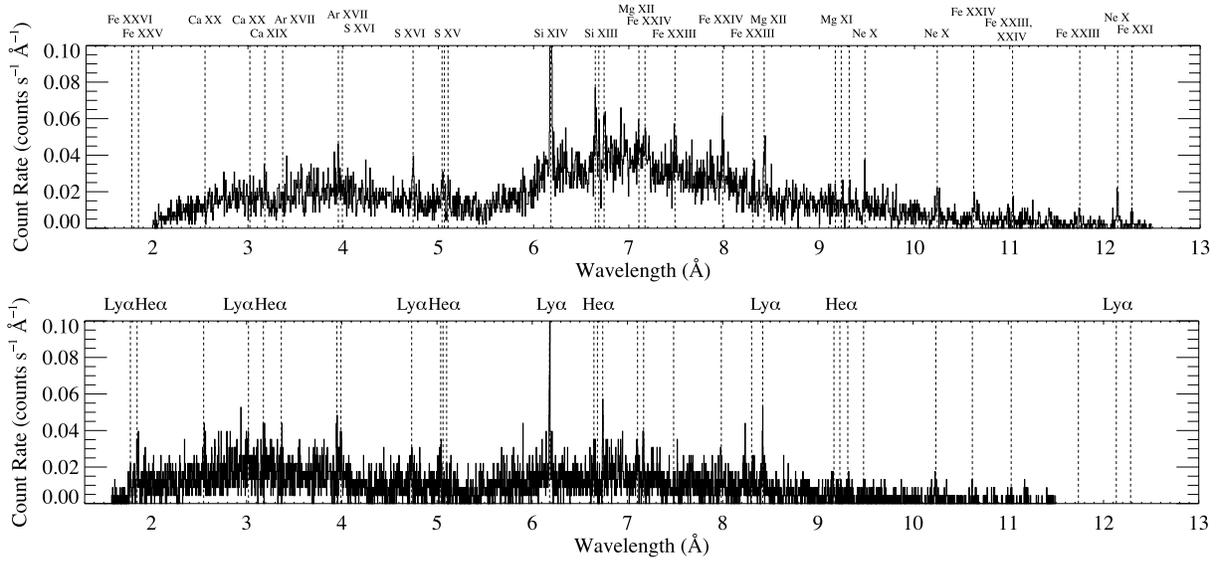}
\caption{The entire usable portions of the MEG (top) and HEG (bottom)
  co-added (negative and positive) first order spectra of DoAr 21.
  The bin sizes are 2.5 m\AA\/ for the HEG and 5 m\AA\/ for the MEG.
  The shape of the continuum is dominated by the wavelength-dependent
  effective area of the telescope, gratings, and detector, and by
  interstellar absorption, rather than by the intrinsic emission.
  Vertical dashed lines represent the laboratory rest wavelengths of
  all detected emission lines. The annotations above the top panel
  indicate the element and ionization stage of each line, while those
  above the lower panel highlight the Lyman-$\alpha$ lines of the
  hydrogen-like isoelectronic sequence and the principal transition of
  the helium-like isoelectronic sequence.  }
\label{fig:bigplot_both}
\end{center}
\end{figure*}

In Figure \ref{fig:bigplot_both} we show the MEG and HEG spectra
(negative and positive first orders co-added in both cases), with
emission lines identified and labeled.  Although the \chandra\/
gratings and detector have significant response to wavelengths above
30 \AA, interstellar attenuation (due to photoelectric absorption,
which has a cross section that goes roughly as $\lambda^3$) makes the
spectra above 12 \AA\/ nearly devoid of counts. These spectra are
dominated by a strong bremsstrahlung continuum, which is indicative of
plasma with a dominant temperature well in excess of 20 million K (so
that atoms are mostly fully stripped and their associated line
emission is weak).  The presence of emission from high ion stages---up
to helium-like Fe\, {\sc xxv} and hydrogen-like Fe\, {\sc xxvi}---also
indicates very high plasma temperatures. We analyze the temperature
distribution in the plasma in detail below by fitting thermal emission
models to the entire spectrum.


\subsection{Spectral Modeling}

We fit the MEG and HEG first order dispersed spectra simultaneously
(but not co-added) over the spectral ranges where each has a
significant number of counts.  For the MEG this was 2--12.5 \AA, and
for the HEG it was 1.5--11.5 \AA.  We fit a two-temperature optically
thin thermal emission model (the {\it bapec} implementation of the
Astrophysical Plasma Emission Code \apec, \citealt{Smith2001}) that
accounts for bremsstrahlung and line emission from a plasma in
statistical equilibrium. This model has four free parameters: the
plasma temperature, the abundances (expressed as a fraction of solar),
the emission measure ($EM = 4 \pi d^2 \int n_{\rm e} n_{\rm H} {\rm d}V$,
proportional to the normalization of the model), and the line
broadening, which is an ad hoc turbulent velocity added in quadrature
to the thermal velocity of each line in the model. We also include
interstellar attenuation, with cross sections from \citet{mm1983}. We
used the $\chi^2$ statistic with Churazov weighting
\citep{Churazov1996} to assess goodness of fit and to place confidence
limits on the derived model parameters.

A single temperature model does not provide a good fit, though the low
resolution \asca\/ and \chandra\/ ACIS data are adequately fit by a
single temperature thermal model \citep{itk2002}.  We did find a good
fit when we used a two temperature {\it bapec} model with interstellar
absorption. The best-fit model has temperatures of roughly 12 and 48
million K (MK), with approximately five times the emission measure in
the hotter component as in the cooler component.  The abundances are
sub-solar, and no significant line broadening is found, with a 68\%
confidence limit of $\sigma_{turb} = 50$ km s$^{-1}$, about one-third
of the peak spectral resolution.  We find an interstellar column
density of slightly more than $10^{22}$ cm$^{-2}$, which is completely
consistent with the extinction of $A_V = 6.2$ magnitudes, given the
conversion between extinction and hydrogen column density of
\citet{Vuong2003}.  The model has a flux of $1.48 \times 10^{-11}$
ergs s$^{-1}$ cm$^{-2}$ on the range 0.3--10 keV. Correcting for
interstellar attenuation, the unabsorbed X-ray flux is a little more
than twice this, corresponding to an X-ray luminosity of $L_{\rm X} =
5.44 \times 10^{31}$ ergs s$^{-1}$.  The best-fit model parameters and
their 68\% confidence limits are listed in the first row of
Table \ref{tab:fit_params}. This fit is formally good, although the
two temperatures certainly are an approximation to a continuous
distribution of temperatures.  Furthermore, the confidence limits are
based only on statistical errors on the data, and in our experience,
for datasets with many bins (the data we fit here have 12,196 bins),
formal confidence limits are unrealistically tight.


\begin{center}
\begin{deluxetable}{lccccccc}
\rotate
\tablecaption{\apec\ model fits\label{tab:fit_params}}
\tablewidth{0pt}
\tablehead{
\colhead{}  &  \colhead{${\rm k}T_{\rm 1}$} & \colhead{$EM_{\rm 1}$}  & \colhead{${\rm k}T_{\rm 2}$} &
  \colhead{$EM_{\rm 2}$} & \colhead{Abundance} & \colhead{$N_{\rm H}(ISM)$} & \colhead{$L_{\rm X}$} \\
\colhead{} &  \colhead{(keV)} & \colhead{($10^{53}$ cm$^{-3}$)} & \colhead{(keV)} & \colhead{($10^{53}$ cm$^{-3}$)}
  &\colhead{(Solar units\tablenotemark{a})} & \colhead{($10^{22}$ cm$^{-2}$)} & \colhead{$10^{31}$ erg s$^{-1}$}}
\startdata
  total & $1.00^{+.04}_{-.03}$ & $6.82^{+.82}_{-.69}$ &  $4.16 \pm .11$ & $30.7^{+0.6}_{-0.7}$ & $0.39 \pm .02$ & $1.19^{+.03}_{-.04}$ & 5.44 \\
  pre-flare &  $0.95^{+.05}_{-.06}$ & $7.40^{+1.37}_{-1.19}$ & $3.10^{+.14}_{-.13}$ & $21.8^{+0.9}_{-1.1}$  & $0.26 \pm .02$ & $1.19^{+.03}_{-.05}$ & 3.49 \\
  flare &  $1.58^{+.08}_{-.10}$ & $18.1^{+4.3}_{-3.6}$ & $7.81^{+2.22}_{-0.81}$ & $40.5^{+1.9}_{-3.3}$  & $0.43^{+.06}_{-.05}$ & $1.18 \pm .05$ & 9.36 \\
  post-flare &  $0.80^{+.09}_{-.05}$ & $11.1^{+2.4}_{-1.6}$ & $4.49^{+.24}_{-.50}$ & $36.9^{+2.1}_{-1.6}$  & $0.31^{+.05}_{-.04}$ & $1.18^{+.06}_{-.05}$ & 6.78 \\
\enddata
\tablenotetext{a}{Solar abundances from \citet{ag1989}.}
\end{deluxetable}
\end{center}

To test the robustness of this fit, we refit the data using the \isis\
X-ray analysis package \citep[ver.\ 1.4.9-38;][]{isis}.  We used the
\apec\ model with interstellar absorption, and, as we did with the
fits in \xspec, found that a two temperature model was required to
achieve a good fit. The temperatures of the best-fit components were
both higher by about 20\%, and the emission measure weighting was even
more skewed toward the hot component in this \isis\ fit compared to
the \xspec\ fit. The interstellar column density was about 10\% lower
than in the \xspec\ fit.  There were some differences in the method
used for the fitting in \isis, including the use of the C statistic
\citep{Cash1979} and a somewhat different wavelength range (1.8--18
\AA\/ for both the MEG and HEG data).  Perhaps these 10--20\%
discrepancies between the results from the two model fitting programs
are a more realistic representation of the parameter uncertainties
than are the formal, statistical confidence limits. We will use the
\apec\ model fitting in \xspec\ as the standard throughout the rest of
the paper.


\begin{figure}
\includegraphics[angle=90,width=170mm]{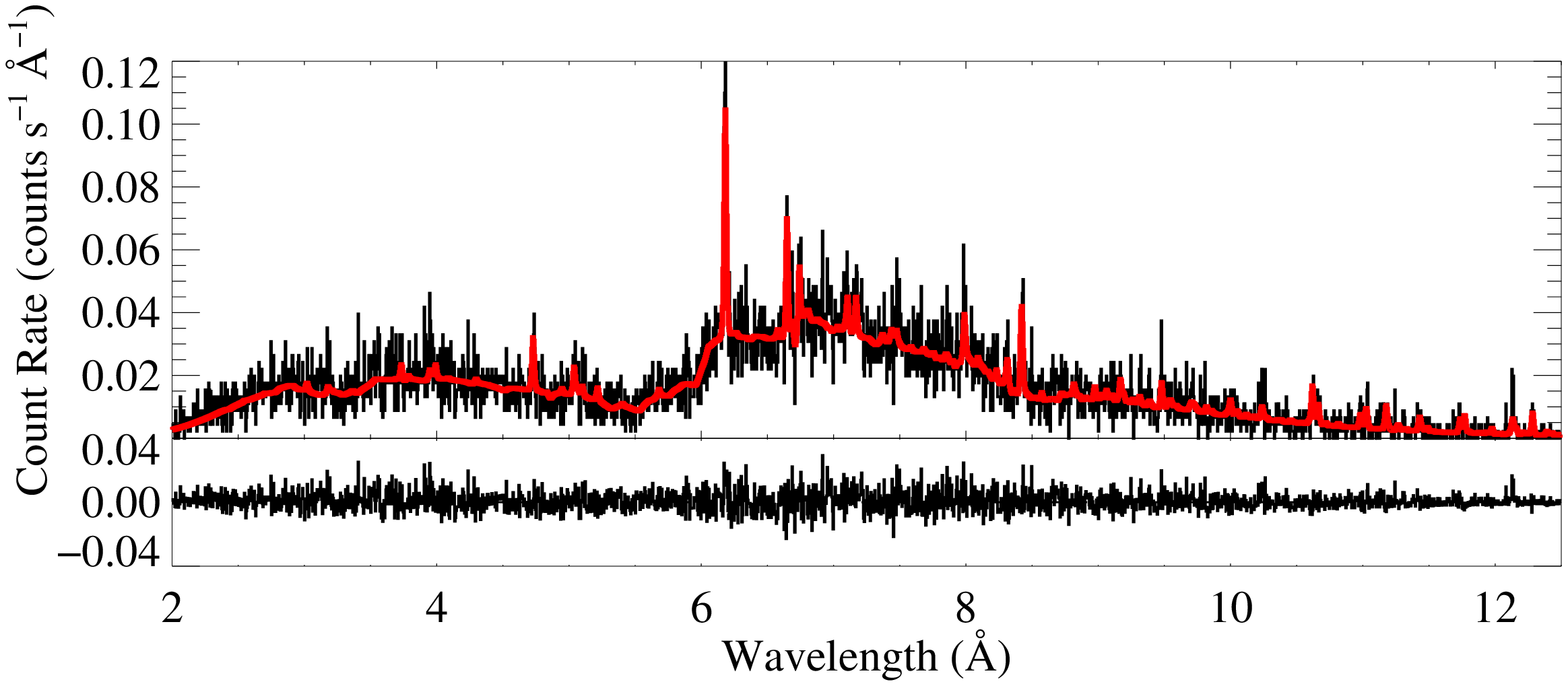}
\caption{The best-fit two-temperature thermal emission model is
  superimposed (in gray; red in the on-line edition) on the \meg\/
  data, with fit residuals shown below.  }
\label{fig:meg_fit}
\end{figure}

The best-fit model to the full observation, with parameters listed in
the top row of Table \ref{tab:fit_params}, reproduces all portions of
the spectrum quite well, with few systematic deviations. The best-fit
model is shown in Figure \ref{fig:meg_fit} along with the higher
signal-to-noise MEG data. In general, the continuum is well fit and
the lines are adequately reproduced by this model. The same is true
for the HEG data.

\subsection{Time Variability and Spectral Analysis of the Pre-flare, Flare, and Post-flare
  Data}

\begin{figure}
\begin{center}
\includegraphics[angle=0,width=80mm]{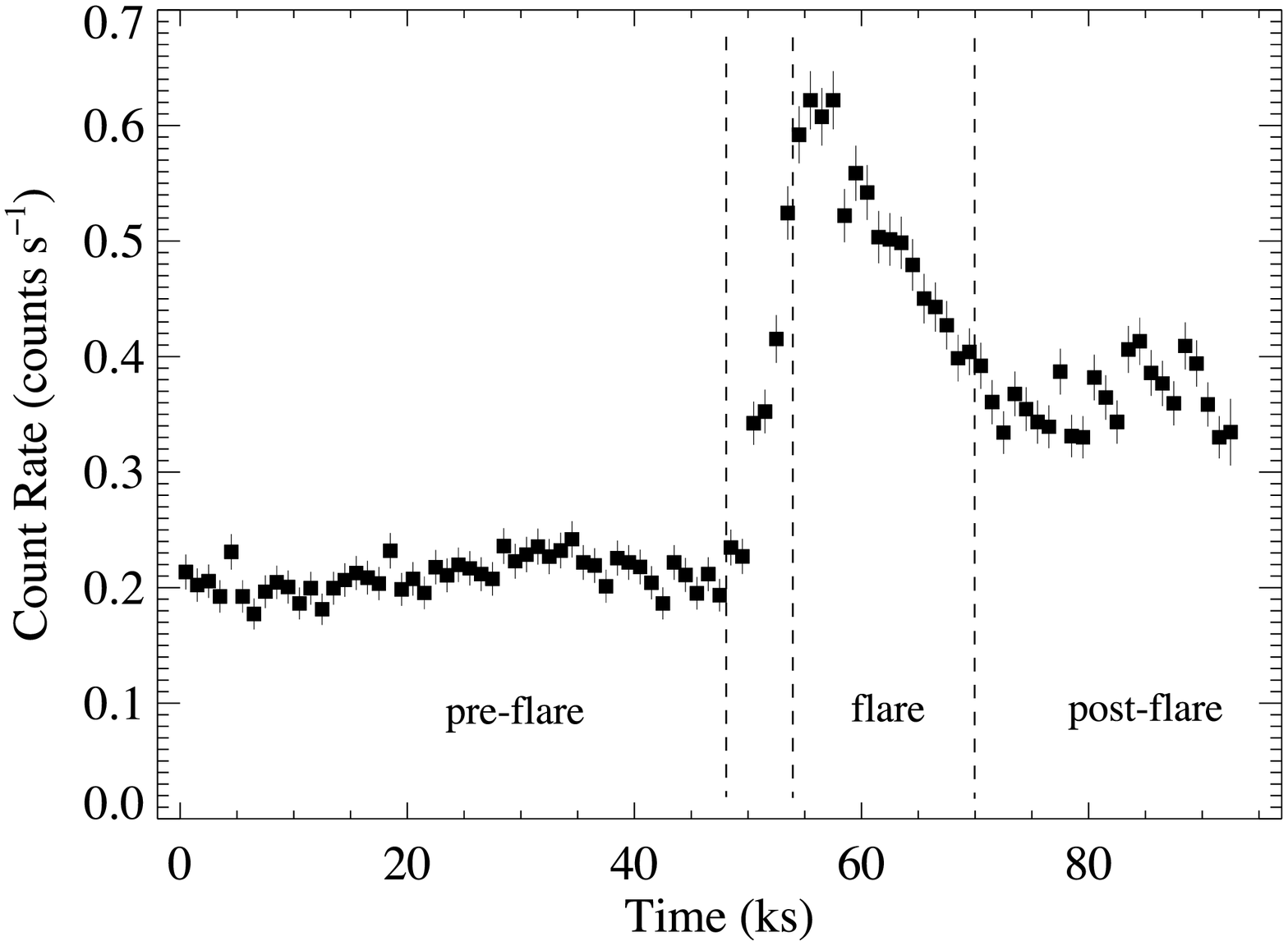}
\includegraphics[angle=0,width=80mm]{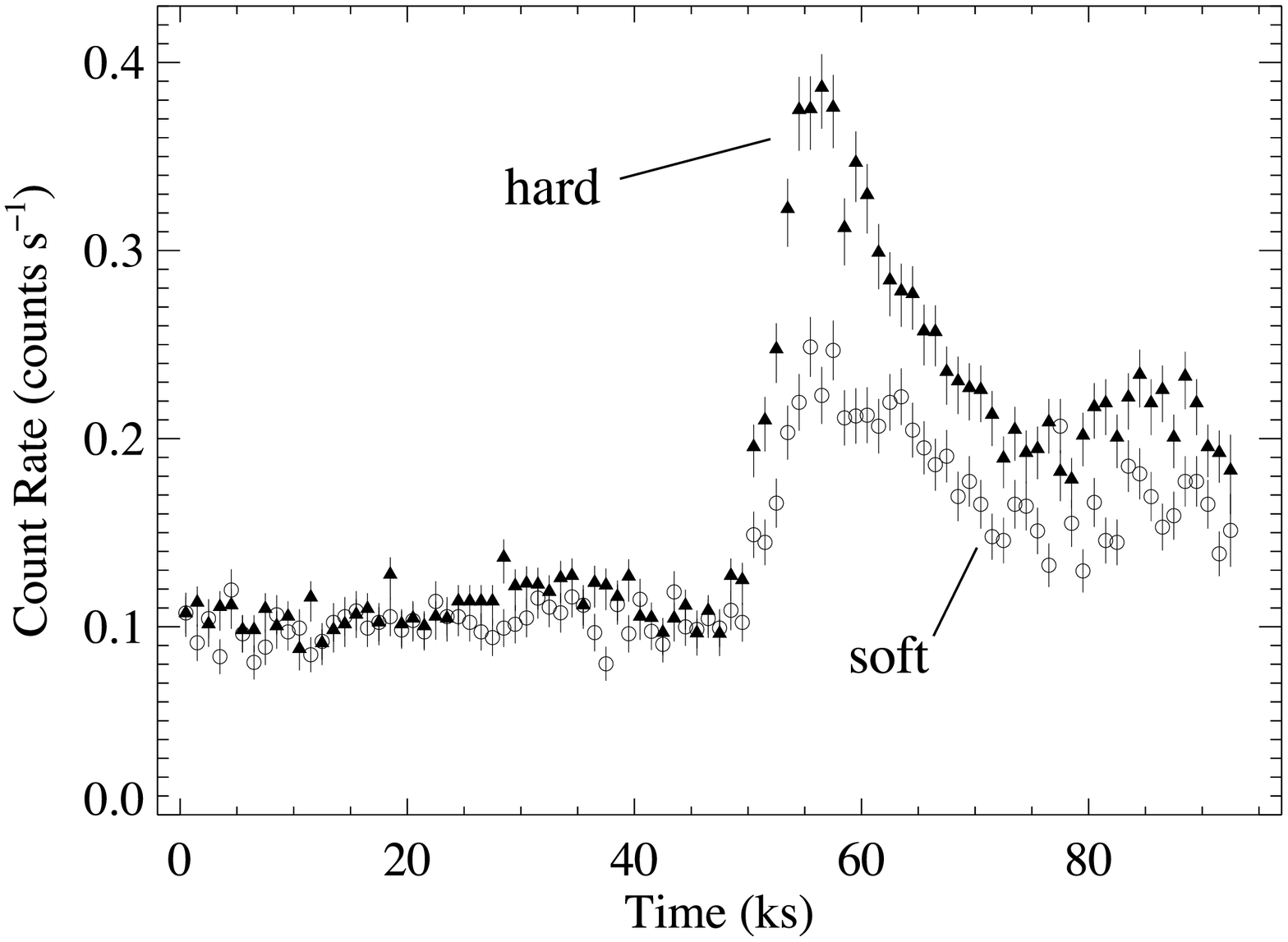}
\caption{Left: A light curve with 1000-second bins formed from all
  counts in the dispersed, first-order spectra (both MEG and HEG).
  Right: Light curves made from all counts with wavelengths longer
  than 6.8 \AA\/ (open circles) and shorter than 6.8 \AA\/ (filled
  triangles). The hardening of the emission during and after the flare
  is evident.}
\label{fig:lightcurve}
\end{center}
\end{figure}

A large (factor of three) flare was seen in the middle of the
\chandra\/ observation, with a rapid rise of about six thousand
seconds, followed by a steady decay during which the count rate drops
by less than a factor of two in about 18 ks.  After this, the count
rate remains steady, but at a level that is nearly twice the
quiescent, pre-flare level. The temporal behavior is shown in Figure
\ref{fig:lightcurve}, where we also display light curves for the hard
and soft bands separately.  Significant hardening is seen during the
flare, which dissipates through the flare and post-flare phase but
never returns to the pre-flare level.


Above we have presented fits to the total spectrum collected during
the 94 ks observation. We next repeat the two-temperature thermal
fitting on the pre-flare (0 through 48 ks), flare (54 to 70 ks), and
post-flare (70 to 94 ks) spectra separately. The results are reported
in Table \ref{tab:fit_params}, and the temporal divisions are
indicated by the vertical lines in the first panel of Figure
\ref{fig:lightcurve}.  The temperatures and emission measures roughly
double during the flare portion of the observation, with somewhat
larger relative changes in the hotter component of the two-temperature
model.  The flare plasma temperature is in excess of 90 MK\null.  The
post-flare plasma also has a temperature distribution that is elevated
with respect to the pre-flare plasma, indicating that some reheating
is occurring.  The absorption column densities derived for each
section of the observation are completely consistent with each other,
which is consistent with an interstellar origin to the attenuation, as
we would expect.  We also note that the abundances show a marginally
significant change among the three different sections of the
observation, in the sense that the hotter sections---the flare and to
a lesser extent the post-flare section---have higher abundances.
Although this result may represent a real change in the coronal plasma
abundances during the flare, it may also be affected by the simplified
two-temperature assumption in the model. The abundance parameter in
the emission model is largely controlled by the line-to-continuum
ratio, and the strength of the metal emission lines relative to the
bremsstrahlung continuum is also affected by details of the emission
measure distribution at high temperatures, where metals are generally
fully ionized and contribute little overall line emission.

Another demonstration of the spectral changes associated with the
flare is provided by the iron K-shell line emission, which is
sensitive to the hottest plasma.  The He-like Fe\, {\sc xxv} line
complex near 1.86 \AA\/ has an emissivity that peaks near 60 MK, while
the H-like Fe\, {\sc xxvi} line at 1.75 \AA\/ has an emissivity that
peaks at temperatures somewhat above 100 MK\null.  There is very
little emission in the He-like feature and none detected in the H-like
feature during the pre-flare section of the observation.  But the iron
lines get significantly brighter during the flare and maintain
significant emission levels during the post-flare phase as well. Not
only do the iron features increase in overall intensity during the
flare, but the relative strength of the H-like feature compared to the
He-like feature also evolves in the sense expected from the
temperature changes seen in the \apec\ model fitting to the entire
spectrum.

\subsection{Density-Sensitive Emission Line Ratios}

The emission lines in the spectrum are relatively weak, but we are
able to analyze the strengths of two important helium-like line
complexes and use them as plasma diagnostics.  Specifically, the
forbidden-to-intercombination line ratios of helium-like ions are
sensitive to density, as collisions de-populate the metastable upper
level $^3$S of the forbidden line and populate the upper level $^3$P
of the intercombination line \citep{gj1969}. Thus higher densities
decrease the forbidden-to-intercombination (\fir) ratio (with
different critical densities for each element).  This evidence for
high densities is seen in some accreting T Tauri stars, but not in
naked T Tauri or magnetically active main-sequence stars \citep[][and
references therein]{Kastner2002, Telleschi2007}.

We measured the line intensities by first fitting the continuum near
each line complex, and once the continuum level was established,
fitting the line complex itself with a three-Gaussian profile model on
top of the best-fit continuum level. The adjustable parameters of the
three-Gaussian model are an overall normalization and the ratios
$\mathcal{G} \equiv ({f+i})/{r}$ and $\mathcal{R} \equiv {f}/{i}$,
where $r$ represents the flux in the resonance line.  The \fir\/
ratios for Si\, {\sc xiii} and S\, {\sc xv} are $5.4^{+2.6}_{-2.2}$
and $1.9^{+2.3}_{-0.8}$, respectively.  None of the lower atomic
number elements have enough signal-to-noise in their helium-like
emission complexes for a measurement to be made.

\begin{landscape}
\begin{figure}
\begin{center}
\includegraphics[angle=90,scale=0.345]{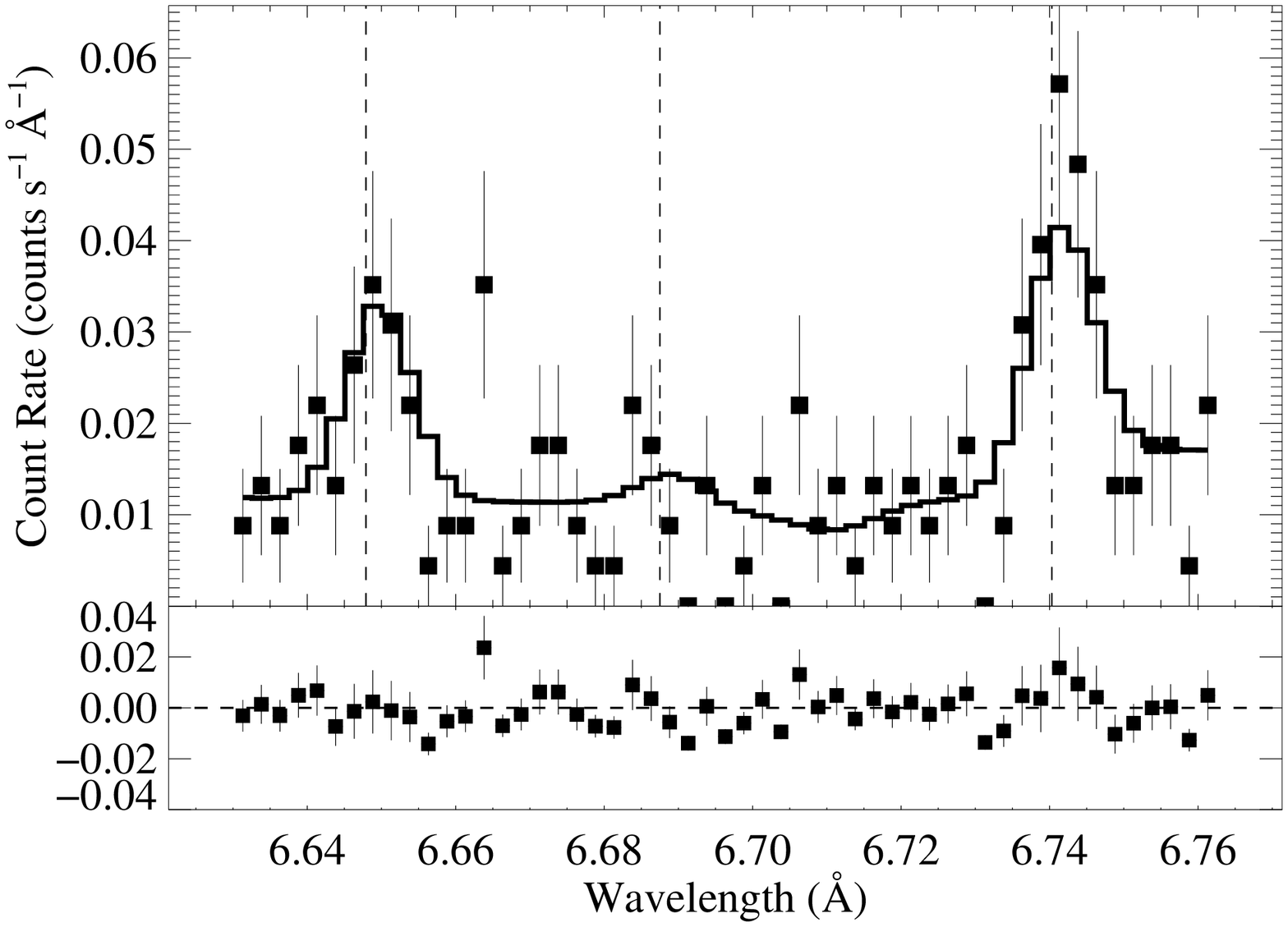}
\includegraphics[angle=90,scale=0.345]{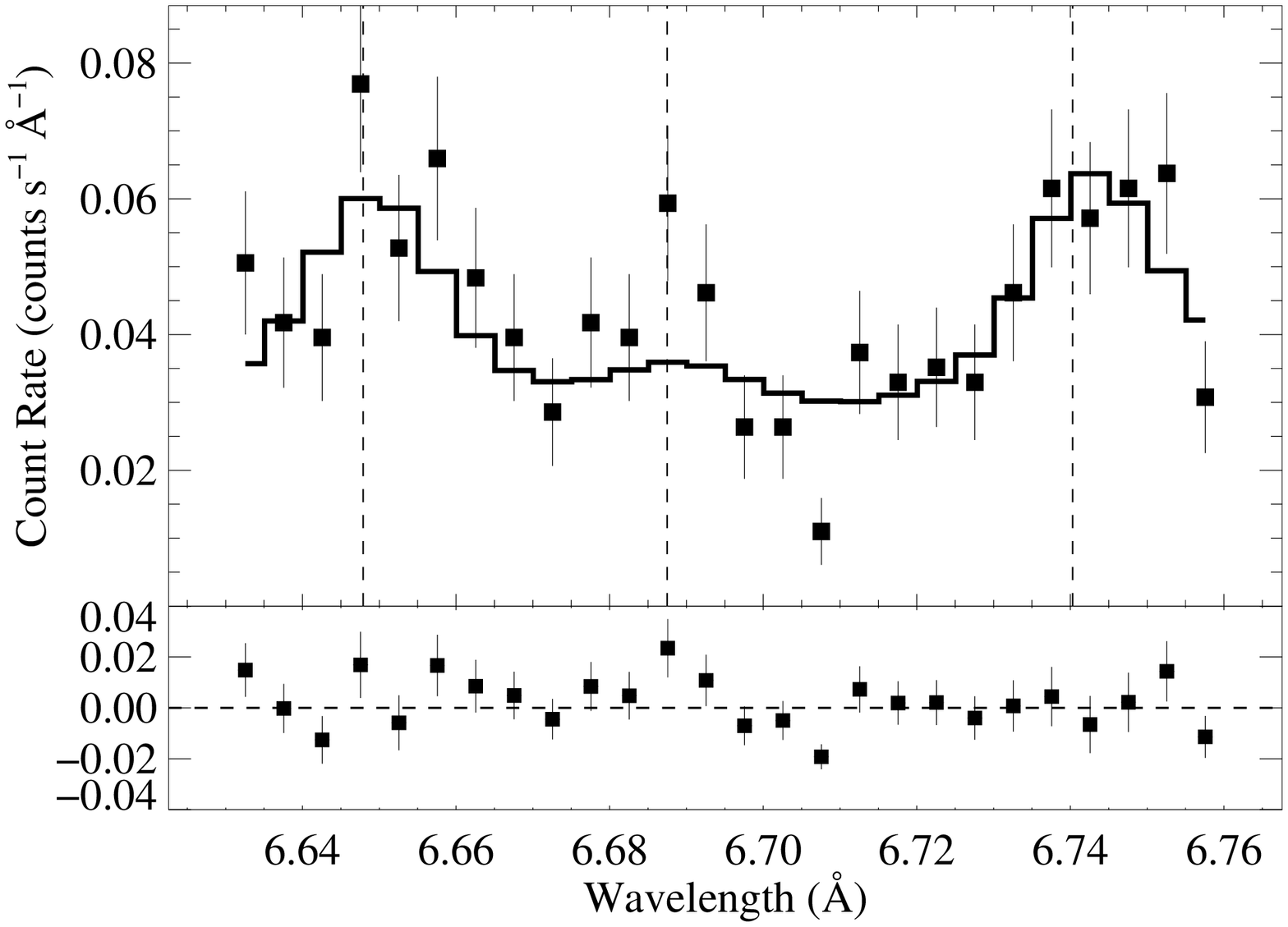}
\includegraphics[angle=0,scale=0.28]{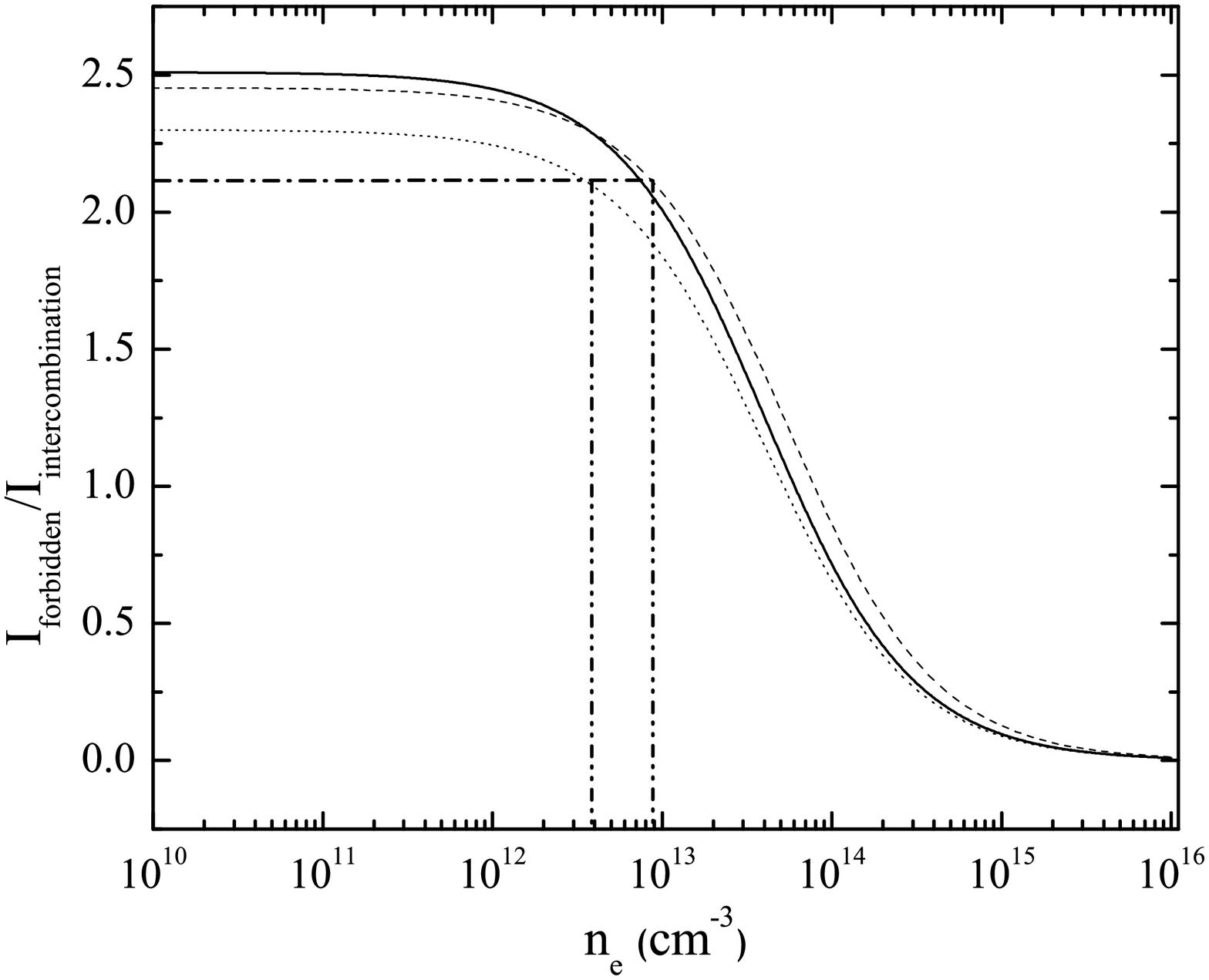}
\end{center}
\caption{The Si\, {\sc xiii} complex in the HEG (left) and MEG (center)
  shows a strong forbidden line (at 6.74 \AA) and a weak
  intercombination line (at 6.68 \AA), consistent with the low-density
  limit. The wavelengths of the three components (resonance,
  intercombination, forbidden from short to long wavelengths) are
  indicated by the dashed vertical lines, and the solid line shows the
  best simultaneous fit to the MEG and HEG data.  Right: Models of
  the \fir\/ ratio of Si\, {\sc xiii} as a function of electron
  density.  The models are from \citet[solid curve]{bdt1972},
  \citet[dotted]{pd2000}, and PrismSpect
  \citep[dashed]{MacFarlane2004}. The horizontal dash-dot line
  represents the 68\% lower confidence limit of $\fir = 2.10$.
  Depending on which of the three models are used, this value
  corresponds to an electron density of $n_e = 4$--$9 \times 10^{12}$
  cm$^{-3}$.}
\label{fig:fir}
\end{figure}
\end{landscape}

For the Si\, {\sc xiii} complex, the forbidden line is blended with
the Lyman-$\beta$ line of Mg\, {\sc xii}.  Based on the measured
Lyman-$\alpha$ flux of Mg\, {\sc xii} and the relative emissivities of
the two components of the Lyman series in the {\sc apec} model, we can
attribute up to 20\% of the measured flux in this feature to the Mg
line.  This makes the ratio $\mathcal{R} \equiv {f}/{i} =
4.3^{+2.6}_{-2.2}$.  The models predict $\fir \le 2.3$--2.5, with the
largest line-ratio values (the ``low-density limit'') expected for all
electron densities less than or equal to a certain limiting
value.\footnote{The low-density limit refers to the limiting (high)
  value of the \fir\/ ratio that holds when the density is low enough
  that collisional excitation out of the metastable excited state of
  the forbidden line is unimportant compared to spontaneous emission
  to the ground state.  We give a range of values for the low-density
  $\fir$ limit for each element.  These represent the range of values
  found from PrismSpect \citep{MacFarlane2004} calculations, from
  \citet{bdt1972}, and from \citet{pd2000}. If the actual electron
  density were much above the low-density limit, then the forbidden
  line would be weakened and the $\fir$ ratio would be reduced.}  The
68\% lower confidence limit on the $f/i$ ratio from our data is 2.1,
which limits the density to less than 4--$9 \times 10^{12}$ cm$^{-3}$,
depending on the model of $\mathcal{R}(n_{\rm e})$ that is employed.
The MEG and HEG measurements of the line complex and the theoretical
models from which we derive the upper limit on the density are shown
in Figure \ref{fig:fir}.


For S\, {\sc xv} the forbidden-to-intercombination ratio we measure is
consistent with the low-density limit for that complex of $\fir= 2.0$
to 2.1. The 68\% lower confidence bound on \fir\/ for S\, {\sc xv}
corresponds to a density of $10^{14}$ cm$^{-3}$ (which represents an
upper limit to the density).  That the helium-like complexes on DoAr
21 have strong forbidden lines, consistent with being in the
low-density limit, is in marked contrast to altered $f/i$ ratios in
accreting classical T Tauri stars (CTTS)\null. However, the altered
$f/i$ ratios in CTTS are generally seen in lower Z elements' He-like
complexes, such as Ne\, {\sc ix} and O\, {\sc vii}, which have lower
critical densities than Si\, {\sc xiii}.  The lower Z features seen in
those CTTS, which are at longer wavelengths, are not visible in the
\chandra\/ spectrum of DoAr 21 because of the large amount of
interstellar attenuation.  Nonetheless, the available density
diagnostics for the X-ray emitting plasma of DoAr 21 are completely
consistent with those seen for active, non-accreting coronal sources
such as weak-lined T Tauri (WTTS) and ZAMS stars.  Specifically, the
upper limit we find for $n_e$ from the Si\, {\sc xiii} complex is
slightly lower than the densities in the accreting T Tauri star TW
Hya, where $n_e = 10^{12.95}$ is found from the Ne\, {\sc ix} complex
\citep{Kastner2002}, and our upper limit is also consistent with
values $\sim 2$ orders of magnitude lower than that, typical of
coronal emission seen in RS CVn stars.

\section{Discussion}
\label{section:discussion}

Our X-ray observations show DoAr 21 to be a very active, hard, coronal
source, with a large X-ray luminosity and flare rate.  At the same
time, our new infrared observations of DoAr 21 show the presence of
extended, irregularly distributed circumstellar material.  In light of
this combination, we now consider the relationship between DoAr 21's
radiation and its circumstellar environment, and what it can tell us
about DoAr 21's evolutionary state and more generally about disk
evolution around young stars.

Much of what we can learn from these data rests on where exactly the
circumstellar material is located, and how it is related to DoAr 21.
Is it the remnant of a disk that is in the final stages of being
cleared out?  Or is it nearby but not necessarily
gravitationally-bound material that is illuminated by DoAr 21 as it
emerges from or passes through the cloud?

We divide the discussion into two major parts.  First, we consider the
emission mechanism and energetics of the observed infrared emission
(including the \molecH\ and PAH emission) from the circumstellar
material; as a corollary to this we evaluate further literature data
bearing on DoAr 21's level of stellar activity, and we use it to
estimate the far-ultraviolet (FUV) radiation field.  We then discuss
possible origins for this circumstellar material, focusing in
particular on whether or not it is a disk, i.e.\ gravitationally
bound, centrifugally supported material.

\subsection{Energetics and excitation of the circumstellar material}
\label{section:energetics}

The most surprising result from the new observations presented here is
the large spatial extent of the mid-infrared emission from DoAr 21,
which is puzzling to find around a cool star, and one with a very low
millimeter flux.  What this discovery has in common with the previous
surprises about DoAr 21---the detections of PAH and \molecH\
emission---is that all of these observations are more commonly
associated with more luminous, hotter stars.  Here we explore whether
it is possible to construct a model that explains all of the data, and
we focus in particular on trying to understand the influence of the UV
and X-ray emission from DoAr 21 on its circumstellar material.
Because FUV radiation is important for radiative
excitation of both \molecH\ and PAH emission, we first discuss the FUV
radiation field near DoAr 21.

\subsubsection{The FUV radiation field near DoAr 21}
\label{section:fuv}

The ultraviolet flux from DoAr 21 is poorly known.  It was not
detected by IUE \citep{Valenti2003}; examination of the archived IUE
spectrum places a rough upper limit of $\sim 10^{-15}$ erg cm$^{-2}$
s$^{-1}$ at $\lambda = 2400$--3200 \AA.  Given the large extinction to
DoAr 21 (roughly 13 magnitudes at $\lambda=2400$ given our adopted
visual extinction), however, the star could have an ultraviolet flux
that is several orders of magnitude above its photospheric flux and it
would not have been detected by IUE\null.  Given this situation, we
take two approaches to estimating the level of UV radiation in the
vicinity of DoAr 21.  First we consider the UV background in the
region due to the nearby B2 star HD 147889.  Then we use the
available data on DoAr 21's chromospheric and coronal activity to
estimate the UV emission from its transition region.

The B2 V star HD 147889, at a projected distance of 0.4 pc from DoAr
21, is responsible for exciting PAH and \molecH\ emission from much of
the surrounding nebulosity \citep{Habart2003}.  To determine the UV
flux from HD 147889 arriving at the position of DoAr 21 and its
surroundings, we integrated the flux from a synthetic spectrum based
on a spherical, hydrostatic, class (ii) NLTE model atmosphere from
\citet{Aufdenberg1999} with $T_{\rm eff} = 22000$ K and $\log g =
3.9$, assuming $L = 5300$ $L_\sun$ for HD 147889 and a distance of 0.4
pc between the two stars \citep{Liseau1999}.  At this distance, the
ultraviolet flux reaching DoAr 21 from HD 147889 (assuming no
attenuation) is roughly 0.75 erg cm$^{-2}$ s$^{-1}$ in the bandpass
6.0--13.6 eV ($\lambda = 912$--2066 \AA).  This corresponds to $G_0
\approx 470$, where $G_0$ is the flux in this wavelength interval
expressed in units of a mean interstellar radiation field of $1.6
\times 10^{-3}$ erg cm$^{-2}$ s$^{-1}$
\citep{Habing1968}\footnote{This is the same quantity referred to as
  $G_{\rm FUV}$ by \citet{NomuraMillar2005} and similar to the
  quantity $\chi$ used by \citet{Habart2003}}. In contrast, the FUV
field from DoAr 21's photosphere (estimated using a Kurucz model
atmosphere and the stellar parameters given above) is $G_0 \approx
220$ at a distance of 10 AU from the star and falls with distance from
DoAr 21.  As noted above, however, given the high level of activity
from DoAr 21, there is likely to be significant UV emission from the
transition region.

To estimate the FUV flux from DoAr 21, we use as a template \hd\ (V987
Tau), a weak-lined T Tauri star with a similar mass, effective
temperature, and X-ray properties.  \hd\ is a single G5 star in the
Taurus-Auriga star-forming region, and it has an X-ray luminosity of
$L_X = 1.1 \times 10^{31}$ erg s$^{-1}$ in the 0.3--10 keV band
\citep[][scaled to the 129 pc distance measured from VLBA parallax by
\citealt{Torres2007}]{Telleschi2007}. DoAr 21's luminosity in this
energy band is $L_X = 3.5$--$9.4 \times 10^{31}$ erg s$^{-1}$ (Table
\ref{tab:fit_params}), taking the range from pre-flare to peak flare
values. \hd\ is near DoAr 21 on the HR diagram, although with a
slightly hotter photospheric temperature and older inferred age; it is
the second blue square from the left in Figure \ref{figure:hr}.  It
has no evidence for circumstellar dust
\citep{Furlan2006,Kundurthy2006} or \molecH\ \citep{Bary2003}.  It
shows a hard, flaring X-ray spectrum \citep{Favata1998}; fitting of
the spectrum yields a temperature distribution similar to what we find
for DoAr 21 \citep{Telleschi2007}.  Like DoAr 21, it shows strong
radio emission \citep{Phillips1991}.

These similarities make \hd\ a good template to compare with DoAr 21,
along with the crucial difference: it has a relatively low $A_V = 0.4$
mag \citep{Furlan2006}.  Thus, the FUV emission is not attenuated
nearly as much as for DoAr 21, and the star was clearly detected by
IUE.  We integrated the de-reddened FUV flux in the co-added \hd\ IUE
spectrum \citep{Valenti2000} in the range $\lambda = 1150$--2066 \AA.
The Ly $\alpha$ line is clearly contaminated by geocoronal emission,
and in any case would be strongly absorbed by interstellar H {\sc I};
we rejected the flux from $\lambda = 1200$--1232 \AA.  Since the
$\lambda = 912$--1150 \AA\ flux was not observed by IUE, we used the
results of \citet{Alexander2005}, who calculated UV spectra using
detailed differential emission measure distributions from
\citet{Brooks2001} for several T Tauri stars and found that the
$\lambda = 912$--1150 \AA\ flux was roughly 1/6 of the $\lambda =
912$--2000 \AA\ flux.  Thus, we scaled the de-reddened IUE 1150--2066
\AA\ flux by 1.2 to estimate the flux in the full 912--2066 \AA\
interval.  We used the measured distances of \hd\ and DoAr 21 (both
with high-quality VLBA parallaxes) to determine the scaling between
flux and luminosity for both stars.  Finally, we scaled the \hd\ FUV
luminosity by the two stars' relative $L_X$ values to estimate the
DoAr 21 FUV luminosity, assuming that the FUV luminosity would be
directly proportional to the X-ray luminosity.

Based on this analysis, we find $L_{\rm FUV} \approx 2.8$--$7.4 \times
10^{31}$ erg s$^{-1}$.  However, this analysis does not account for
flux in the Ly $\alpha$ line, which can be very strong in stars with
active chromospheres.  We estimated the \lya\ flux using the scaling
relations between \lya\ and X-ray surface flux derived by
\citet{Wood2005}, who measured \lya\ from a sample of nearby late-type
stars at high spectral resolution with STIS on HST, and thus were able
to remove geocoronal emission and correct for interstellar absorption.
The range of X-ray fluxes from DoAr 21 gives a range of \lya\ surface
fluxes of 0.8--$1.3 \times 10^7$ erg cm$^{-2}$ s$^{-1}$,
and thus a total $L_{\rm FUV} \approx 3.6$--$8.7 \times 10^{31}$ erg
s$^{-1}$.  We caution, however, that there is only one star (Speedy
Mic) in the \citet{Wood2005} sample with an X-ray surface flux as
large as that from DoAr 21, and so the scaling relations are not
well-constrained in this regime.

Under these assumptions, we estimate that the $G_0$ value produced by
DoAr 21's FUV flux is in the range of 470--1100 at 1\farcs1 (134 AU)
from the star, the distance of the bright ridge in our images.  Thus,
the estimated FUV flux from the star is equal to or somewhat greater
than the external FUV flux at this distance.  The total FUV flux (from
the combination of DoAr 21's transition region and the B2 star
background) at the position of the bright ridge is then $G_0 \approx
940$--1600; the internal and external FUV fluxes become equal at
$\sim$ 130--210 AU from DoAr 21.  If there is any attenuation of the
external FUV flux in the $\ge 0.4$ pc between HD 147889 and DoAr 21,
then DoAr 21's contribution to the total flux is even more dominant. 

Having estimated the FUV flux near DoAr 21, we can now examine the
effect this flux would have on nearby material.  In the sections
below, we will examine in turn the continuum (dust) emission, the PAH
emission, and the \molecH\ emission.  Consistent with the FUV flux
estimates above, we will show that the emission shows signatures of
being excited from the inside out, i.e.\ by DoAr 21 rather than
externally by HD 147889.  Before turning to a discussion of the
details of each type of emission, we close this section with a simple
check of the energy balance in the circumstellar environment.  We
measured the flux from the circumstellar material by taking the fluxes
in the largest aperture listed in Table \ref{table:gemini-photometry}
and subtracting the fluxes in the smallest aperture to remove the
stellar contribution.  Integrating from 10 \micron\ to 20 \micron, we
find the luminosity of this material to be roughly $9 \times 10^{31}$
erg s$^{-1}$, the same order of magnitude as our estimate of $L_{\rm
  FUV}$.  Given the uncertainties in our flux estimate and the
complexities of the various heating, cooling, and emission processes
in the environment we do not suggest that there should be precise
agreement here, but the order-of-magnitude agreement does indicate
that DoAr 21 is a plausible source for powering the observed emission,
which we now examine in more detail.

\subsubsection{Dust emission}

In order to study the continuum emission excess from DoAr 21, which we
here assume arises from thermal emission from dust grains, we need
wavelength regions without significant spectral features.  The
spectrum taken by \citet{Hanner1995} shows that the region encompassed
by the 10.4 \micron\ filter from our Gemini observations is free of
emission or absorption features; in particular, there is no evidence
of a silicate feature.  While we do not have a corresponding spectrum
in the 18.3 \micron\ region, this part of the spectrum generally has
not shown significant features in Spitzer IRS spectra of other young
stars \citep{Geers2006}.  Some galaxies and star-forming regions with
PAH emission do show features at 17.8 \micron\ and 18.9 \micron,
albeit on top of a significant continuum \citep{Sellgren2007}.  We
take this spectral region to sample primarily continuum emission; if
there is a contribution from some PAH emission in this bandpass, the
effect would be to raise the dust temperatures derived below.

Using our Gemini images, we integrated the flux in these two
bandpasses in a region that is off the stellar position and thus
probes only circumstellar flux.  This yielded fluxes of 197 and 603
mJy at 10.4 and 18.3 \micron, respectively.  A single-temperature
blackbody fit to these two fluxes yields a temperature estimate of
roughly 215 K for the emitting material.  While this estimate is
obviously an oversimplification of the actual situation, the general
result of a steeply-rising flux from 10--18 \micron\ is robust,
indicating that the emitting material has a temperature in the
hundreds of degrees K.  For comparison, blackbody grains that have
reached thermal equilibrium with the star's radiation field, given
DoAr 21's luminosity of 11.7 $L_\sun$, would have a temperature of 44
K at 135 AU (1\farcs1) from the star and 31 K at 270 AU from the star.
At these temperatures, the 18.3 \micron\ to 10.4 \micron\ flux ratio
would be 10$^5$--10$^7$, rather than the observed value of $\sim 3$.

While the derived dust temperature is far out of equilibrium, it is
quite consistent with the expected behavior for very small grains
(VSGs) excited by UV photons \citep{PugetLeger1989}.  Grains with
radii of a few nm can be transiently heated to temperatures of several
hundred K by single UV photons, and an ensemble of such grains can
contribute significant infrared emission.  Given that such small
grains may form a continuum of sizes with PAH molecules
\citep{Tielens2008}, it is perhaps not surprising to find them around
DoAr 21.  Nevertheless, it provides additional evidence for the
importance of the FUV radiation in exciting the circumstellar
material, and it is another respect in which DoAr 21's environment is
different from a typical circumstellar disk.

\subsubsection{PAH emission}
\label{section:pah}

We have presented our new images in the 8.6 and 11.3 \micron\ PAH
filters above; here we briefly recap existing published and archival
PAH data for DoAr 21.

Figure \ref{figure:pah} shows the mid-infrared spectrum of DoAr 21
from \citet{Hanner1995} along with the model photosphere. The data
show a clear emission feature at $\lambda = 11.3$ \micron, attributed
by \citet{Hanner1995} to emission from PAHs.  The data also show a
rising slope shortward from $\lambda = 8.5$ \micron\ to where the
spectrum ends at 7.9 \micron, consistent with the presence of a 7.7
\micron\ PAH emission feature.

DoAr 21 was also observed by ISO with the circularly-variable filter
(CVF) on ISOCAM, yielding a $\lambda/\Delta \lambda \approx 50$--80
spectrum from 5.1--16.6 \micron\ of DoAr 21 and its surroundings with
6\arcsec\ pixels.  These data were reprocessed by
\citet{Boulanger2005} as part of the final reprocessing of all ISOCAM
CVF data, and we retrieved the reprocessed data from the ISO archive.
The data show a strong point source at the position of DoAr 21, as
well as a bright ridge of emission located 30\arcsec--90\arcsec\ NE of
DoAr 21 and running NW--SE; this ridge is clearly visible in the
\spitzer\ 24-\micron\ image (Fig.\ \ref{figure:spitzer-image}; see
also Fig.\ 3 of \citealt{Padgett2008} for a larger view) and can also
be seen faintly in the 2MASS K-band images.  All pixels in the ISOCAM
3\farcm3 x 3\farcm3 field of view show a rich PAH spectrum, including
emission features at 6.2, 7.7, 8.6, 11.3, and 12.7 \micron.

Unfortunately, it is not possible to determine a reliable spectrum for
DoAr 21 itself from these data.  Due to internal reflections in ISOCAM
when used with the CVF, the spectrum of DoAr 21 is contaminated by
``ghosts'', reflections of stray light from elsewhere in the field of
view \citep{Okumura1998}.  This is revealed by sharp discontinuities
in the spectrum around 9 \micron, the transition wavelength between
CVF1 and CVF2.  The ghosts are significantly worse in CVF2, and this
discontinuity (which is clearly present in a spectrum extracted at the
position of DoAr 21) is a signature of stray light contamination
\citep{Boulanger2005}.  We suspect that the contamination arises from
the bright ridge of emission to the NE of DoAr 21, which is of the
same order of magnitude in brightness at these wavelengths as DoAr 21
itself.  Because the background emission is diffuse and spatially
variable, and the ghost size and strength are functions of wavelength,
it is not possible to reconstruct the uncontaminated spectrum of DoAr
21 from these data.

One of the most striking features of our new PAH-filter images of DoAr
21 is the spatial variation of the 11.3 \micron\ emission, with
similar variations seen at 18.3 \micron.  The emission does not
decrease smoothly with distance in the region near the star, but
rather it is significantly brighter in a knot and partial arc
beginning roughly 1\farcs1 to the north and west of the star.  In
contrast, the 8.6 \micron\ emission is much smoother, with about the
same brightness in the region near the star as in the region that
appears as a bright arc in the 11.3 \micron\ image.  Thus, it appears
that the [8.6]/[11.3] ratio may be higher near the star.  However, the
8.6 \micron\ emission is much fainter overall than the 11.3 \micron\
emission, and the star is brighter, so it may be that any variation in
the fainter 8.6 and 10.4 \micron\ emission blends in with the wings of
the stellar PSF.

If the apparent change in the [8.6]/[11.3] ratio is real, it is a
useful diagnostic of the ionization state of PAH molecules; in
laboratory spectra of PAH molecules, the 11.3 \micron\ CH out-of-plane
bending mode weakens considerably in ionized PAHs while the 8.6
\micron\ CH in-plane bending mode strengthens somewhat
\citep{Allamandola1999}.  In astrophysical sources, a decrease in the
[8.6]/[11.3] ratio with distance from a central ionizing source has
been observed in NGC 1333 \citep{Joblin1996} and in the Orion Bar
\citep{Galliano2008}.

Applying this interpretation to DoAr 21 leads to the conclusion that a
significant fraction of the PAH molecules within about 1\arcsec\ of the
star are ionized, suppressing the [8.6]/[11.3] ratio in this region.
This would indicate that DoAr 21, not HD 147889, is the dominant
source of FUV flux in its circumstellar environment, consistent with
our estimates above of the relative FUV fluxes (Sec.
\ref{section:fuv}).  The first ionization potential for typical PAH
molecules is 6--8 eV \citep{wd01b}, and thus there must be significant
flux at energies $\gtrsim 6$ eV near the star.

As noted above, it may be that the radial variation seen in the bright
11.3 \micron\ emission is simply difficult to detect in the fainter
8.6 \micron\ emission due to the low surface brightness.  One piece of
evidence in favor of this interpretation comes from comparison of the
different continuum wavelength bands.  Since both the 10.4 and 18.3
\micron\ bands are dominated by continuum emission from very small
grains, they should show the same spatial variation.  However, the
bright 18.3 \micron\ emission shows the same radial variation as the
bright 11.3 \micron\ emission, while the fainter 10.4 \micron\
emission apparently does not, which might suggest that it is not seen
simply due to low surface brightness, and that a similar effect might
be present at 8.6 \micron.  If it is the case that the same radial
variation is present in all bands, then it would represent a real
change in surface density of material, with less material present near
DoAr 21, perhaps due to photoevaporation.  In either case, the
azimuthal distribution of material around DoAr 21 is clearly
asymmetric (Fig.  \ref{figure:gemini-azimuthal-profile}).

While there is a sharp increase in the 11.3 \micron\ emission at
roughly 1\farcs1 from the star, the brightness of this emission
declines with radial distance beyond that point.  The decline is
roughly consistent with a falloff proportional $r^{-2}$, which is what
would be expected from a uniform PAH and small grain surface density
with declining excitation due to geometric dilution of the exciting
radiation.  Thus, the true extent of the circumstellar material may be
much greater than that shown in our images, as already suggested by
the HST and near-infrared images (Sec.\ \ref{section:gemini-data}).

\subsubsection{H$_{\rm2}$ emission}

While we did not make any observations of \molecH\ emission, the
analysis of the VSG and PAH emission in the preceding sections bears
directly on the interpretation of previous \molecH\ observations,
which we discuss here.  Given the surprisingly large
spatial extent of the emission seen in our images, we pay particular
attention to what constraints existing data put on the spatial
distribution of the \molecH\ emission.  

As noted above, \citet{Bary2002,Bary2003} detected emission in the
\molecH\ \Sten\ line at $\lambda = 2.12$ \micron\ from DoAr 21.  Their
slit width was 0\farcs8 and the seeing was approximately 1\farcs4
\citep{Bary2002}. In their nod-differenced images, they found no
evidence of any extension of the emission along the E-W slit beyond
the 1\farcs4 seeing disk.  From the resolved velocity width of the
line, they also argued that the gas must reside in a rotating
circumstellar disk.

\citet{Bitner2008} observed DoAr 21 in the 0--0 S(2) pure rotational
transition of \molecH\ at $\lambda = 12.279$ \micron.  They detected
emission in one observation using a 1\farcs4 N-S slit with 1\farcs4
seeing on the IRTF, but they did not detect it in a subsequent
observation with a 0\farcs54 E-W slit with 0\farcs5 seeing on
Gemini.\footnote{FWHM seeing profiles and slit orientations are from
  M. Bitner, personal communication.  The stated slit orientations are
  the direction of the long dimension of the slit, i.e.\ the direction
  in which any extended emission would be detected.}

The \citet{Bitner2008} observations suggest that the S(2) \molecH\
emission is not uniformly distributed around the star and/or is
time-variable.  If we attribute the difference between the two
observations solely to the spatial distribution of material, the data
imply that the emitting gas lies primarily at an angular separation
greater than 0\farcs25 (30 AU), and that it may be more concentrated
to the N and/or S of the star.  \citet{Bitner2008} detected larger
fluxes with the wider IRTF slit than with the narrower Gemini slit for
four out of the five sources observed with both telescopes, suggesting
that spatially-extended emission for the pure rotational transitions
is not uncommon.  We note, however, that a slit at any orientation
will intercept flux from a disk at all disk radii (though not all
azimuths), so azimuthally-symmetric emission from gas in a disk, even
at large radii, should contribute some flux even in the Gemini slit.
We discuss this point further in Sec.\ \ref{section:disk}, where we
propose a possible geometry for the emitting material that is
consistent with all the data.

The suggestion that the \molecH\ S(2) emission may lie preferentially
to the N or S, combined with the emission geometry seen in our PAH
images, suggests that the \molecH\ and PAH emission could be
coincident, and other observations suggest that this is not an
uncommon occurrence. This is perhaps unsurprising, given that both
molecules are excited by UV photons.  In the Orion Bar PDR, the
\molecH\ \Sten\ emission and 3.3 \micron\ PAH emission are similarly
distributed along the edge of the bar facing the ionizing O
stars.\footnote{While to our knowledge there is not a similar map of
  the 11.3 \micron\ line in the Orion Bar, the 3.3 \micron\ PAH line
  is strongest in neutral PAHs and in general is strongly correlated
  with the 11.3 \micron\ line \citep{Hony2001}.}  Closer to DoAr 21,
\citet{Gomez2003} and \citet{Habart2003} conducted ground-based
narrow-band imaging, showing that the SW edge of the bright bar lying
NE of DoAr 21 (seen in both the \spitzer\ and ISO images) is also
bright in the \molecH\ \Sten\ line.  Further, \citet{Habart2003} show
that the PAH emission seen by ISOCAM at $\lambda=5$--8.5 \micron\ in
this region correlates very well in strength and position (albeit with
modest spatial resolution) with the \molecH\ \Sten\ emission seen in
the ground-based images, and with the \molecH\ 0--0 S(3) emission seen
by the ISO SWS\null.

An alternate interpretation of the \citet{Bitner2008} observations is
that the emission is time-variable and was below the detection
threshold at the time of the Gemini observation.  The $3\sigma$ upper
limit in the Gemini observation is $1.25 \times 10^{-15}$ erg
cm$^{-2}$ s$^{-1}$, a factor of 2.6 below the flux level of the IRTF
detection.\footnote{This limit is determined using the line width of
  5.6 km s$^{-1}$ measured in the IRTF observations (M. Bitner,
  personal communication), and thus it is slightly lower than the
  limit given in \citet{Bitner2008} based on an assumed line width of
  10 km s$^{-1}$.}  The X-ray flux level from DoAr 21 is seen to vary
by a factor of three on short timescales (Fig.\ \ref{fig:lightcurve});
if this translates into a corresponding variation in the FUV exciting
the \molecH\ emission, then that emission could vary by similar
factors.  However, if we attribute the difference between the two
observations solely to variability, this hypothesis would require that
the \citet{Bitner2008} detection happened to coincide with the peak of
a flare, and that the \molecH\ flux in the Gemini observation is just
below the detection threshold.  DoAr 21 has a relatively high X-ray
flare rate of about one per day \citep{itk2002}, but the fraction of
time the X-ray flux is more than a factor of 2.6 above the quiescent
level is small, probably less than an hour per day based on available
data.

\subsubsection{Photometric and Polarimetric Variability}
\label{section:opt-variability}

Multiple measurements exist at wavelengths from 0.39 \micron\ ($U$)
through 4.8 \micron\ ($M$).  Variability is seen at all wavelengths.
The range at $V$ band in 15 measurements is 0.33 magnitudes, with
variations on timescales of less than one day, but with no apparent
periodicity \citep{Bouvier1988}.  Substantial variability is seen in
the ultraviolet; in two observations separated by less than an hour,
\cite{Bouvier1988} observed a brightening of 0.97 magnitudes in $U$,
while $B$ brightened by 0.03 magnitudes and $V$ faded by 0.02
magnitudes.  Two observations separated by 3 hours on the following
night showed a similar variation, with $U$ brightening by 0.42
magnitudes while $B$ and $V$ faded by 0.04 and 0.05 magnitudes,
respectively.\footnote{\citet{Bouvier1988} do not give error bars on
  individual measurements.  They quote typical errors of 0.09, 0.06,
  and 0.04 mags for $U$, $B$, and $V$ for a $V=12$ star; DoAr 21 is
  $V\approx 14$.  Observations of other stars with similar $V$
  magnitudes observed during the same nights as the $U$ flares on DoAr
  21 do not show $U$ magnitudes that are substantially different from
  the mean values for those stars, suggesting that the observed
  changes at $U$ in DoAr 21 are real, and not photometric errors.}
Such a large increase in the UV emission without a concurrent increase
in $V$ cannot be due to rotational modulation (either of hot or cool
spots), since that would increase the flux in all bands
\citep[e.g.,][]{RydgrenVrba1983}.  These $U$-band brightenings are
similar to flares seen on active main-sequence stars in that they are
much brighter at $U$ than at $V$ \citep{Fernandez2004}.  We note that
both of these brightening events occur when the star is at the faint
end of the observed $U$ range, i.e.\ they are brightenings toward the
mean value rather than above it.  This may simply indicate that the
star is flaring much of the time and that the observed mean value in
the data of \citet{Bouvier1988} is characteristic of the flares rather
than the photosphere; this is consistent with the fit to the SED
(Fig.\ \ref{figure:sed}).  The $U$ and $B$ magnitudes are positively
correlated with each other (with a larger amplitude at $U$), but
neither the $U$ or $B$ magnitudes nor the $U - B$ color is correlated
with $V$.

The bluest observed $U-B$ color for DoAr 21 in the photometry of
\citet{Bouvier1988} is $U-B = 0.5$, observed on two consecutive
nights.  Using our adopted extinction, the de-reddened color is
$(U-B)_0 \approx -0.6$.  Interestingly, the only other stars in the
\citet{Bouvier1988} sample that show $U-B$ excursions to such blue
values are classical T Tauri stars, though the availability of
well-sampled time series $U$ and $B$ photometry for some of the most
active WTTS (e.g., \hd) is limited.  On DoAr 21, where we see no
evidence of accretion, the $U$-band excess probably signals
Balmer-line and continuum emission resulting from extreme
chromospheric activity and flaring, as seen on late-type flare stars
\citep[e.g.,][]{Lacy1976}.  Flares from such stars have a mean $U-B$
of $-0.9 \pm 0.3$ \citep{Moffett1974}, consistent with the bluest
color seen from DoAr 21.

DoAr 21's $K$-band emission is polarized at $p \approx 2$\%
\citep{Martin1992}.  The large visual extinction toward DoAr 21
suggests that this could be attributed to interstellar polarization
from overlying cloud material not closely associated with the star.
However, the polarization is variable both in amplitude and in
position angle \citep{Jensen2004}.  Since a cloud far in the
foreground would not be expected to change on short timescales, this
variability suggests a circumstellar origin for DoAr 21's
polarization; this is also quite consistent with the observed
asymmetry of the circumstellar material, as scattering off an
asymmetric distribution will give a net polarization in the unresolved
light.  Given the discovery reported here that the circumstellar
material around DoAr 21 is quite extended (Sec.\
\ref{section:gemini-data}), the observed variation in polarization
percentage could be due to measuring with different aperture sizes
that include different amounts of the extended emission: the
percentage polarization measured by \citet{Jensen2004} in a
4\farcs3-diameter aperture is smaller than that measured by
\citet{Martin1992} in a 7\farcs8 aperture.

\subsection{A Disk or Not?}
\label{section:disk}

What is strikingly clear from the observations presented above is that
the high-energy emission from DoAr 21 has strong interactions with
material in its circumstellar environment, even on scales of hundreds
of AU.  What is less clear, however, is the relationship of the
observed material to DoAr 21, and specifically to what extent the
extended emission is related to any orbiting, bound circumstellar (or
circumbinary) disk that DoAr 21 has or once had.

The definitive answer to this will await further, higher-resolution
observations, in particular spatially- and velocity-resolved mapping
of the \molecH\ emission.  Here, we close the discussion with an
examination of evidence for and against a Keplerian disk around DoAr
21.

One of the primary arguments that \citet{Bary2002} made for the
\molecH\ emission arising from a disk is that extended, non-disk
material would have a narrower line width than the 9 \kms\ FWHM width
they observed for DoAr 21's \Sten\ emission.  While we examine this
line width more quantitatively below, we first note that DoAr 21 has
the narrowest \molecH\ line of any of the detections of \molecH\ from
T Tauri stars and Herbig Ae/Be stars at both \Sten\
\citep{Bary2003,Itoh2003,Takami2004,RamsayHowat2007,Bary2008,Beck2008}
and 0--0 S(2) \citep{Bitner2008}.

To interpret the line width, we must take into account the fact that
the observed line width is the convolution of the intrinsic line width
with the instrumental profile.  The \citet{Bary2002} observations had
a FWHM spectral resolution of 5 \kms.  Assuming Gaussian profiles for
both the spectral response of the instrument and the observed line,
the intrinsic line width is then 7.5 \kms.  If interpreted solely as a
thermal line width, this would imply a kinetic temperature of 2450
K\null.  \citet{Bary2002} do not quote an uncertainty for the measured
line width; assuming a 1 \kms\ total uncertainty for the deconvolved
line width (reflecting some uncertainty in both the measured line
width and the spectral resolution), the range of derived kinetic
temperatures is then 1700--3300 K\null.  The lower end of this
temperature range is quite consistent with temperatures generally
found for \Sten\ emitting gas.  For example, \citet{Bary2008} quote a
standard thermal excitation temperature of 2000 K\null.
\citet{Beck2008} observed extended \molecH\ around several T Tauri
stars, and from the ratios of several lines they derive LTE
temperatures of 1800--2300 K for \molecH\ around T Tauri stars.
\citet{Herczeg2004} model the fluorescent \molecH\ emission from TW
Hya's disk and find kinetic temperatures of $2500^{+700}_{-500}$ K for
the \molecH.  While the gas in some of those sources may be
shock-heated, which is unlikely for DoAr 21, these observations show
that \molecH\ can survive and emit strongly at the temperatures
necessary to explain the observed line width in DoAr 21.  Similarly,
the models of \citet{NomuraMillar2005} and \citet{Nomura2007} predict
that the bulk of the \molecH\ \Sten\ emission from disks irradiated by
X-rays and UV comes from gas with temperatures of 1000--2000 K, and
similar results are found in models of PDRs \citep{Allers2005}.
Collisional dissociation of \molecH\ at these temperatures is
insignificant \citep{Martin1996}.  Thus, the \molecH\ temperature
necessary to explain DoAr 21's observed line width solely with thermal
broadening is consistent both with theoretical expectations and with
observations of other sources.

Theoretical models show that it is plausible that DoAr 21 can heat gas
to these temperatures at large distances from the star.  The FUV flux
that we estimate for DoAr 21 combined with HD 147889 (Sec.\
\ref{section:fuv}) corresponds to $G_0 \approx 940$--1600 at a
distance of 130 AU (1\farcs1) from DoAr 21, the location of the bright
PAH emission ridge.  This is similar to the FUV flux in the
\citet{NomuraMillar2005} models at radii of $ \sim 40$--100 AU, and
thus we might expect conditions at 130 AU from DoAr 21 to be similar
to those at 40--100 AU in the models.  This range of radii extends
beyond the $\sim 30$ AU distance at which the \molecH\ \Sten\ emission
peaks in those models, but it is also the case that DoAr 21 has an
X-ray luminosity that is 50 times larger and an X-ray spectrum that is
much harder than that assumed in these models (which base their
assumed X-ray emission on that of TW Hya), thus contributing an
additional source of heating.  Thus, while the models do not apply
directly to DoAr 21, plausible scaling from them suggests that
\molecH\ at $ > 100$ AU from DoAr 21 could have $T > 1000$ K,
necessary to produce the observed line width with thermal motions, and
could emit at the observed level, since the predicted \molecH\ flux in
both the \Sten\ and pure rotational lines in these models is similar
to that observed for DoAr 21.

To summarize this part of the discussion, the primary argument of
\citet{Bary2002} for the presence of a disk around DoAr 21 is
essentially that the only way to produce the observed \molecH\ \Sten\
line width is from rotation in a disk.  However, we have shown that
the observed width is a plausible thermal line width for \molecH\
illuminated with a strong FUV and X-ray flux, and that the flux from
DoAr 21 is sufficient to produce the necessary heating at distances of
$> 100$  AU from the star, consistent with the position of the observed
PAH and VSG emission.  While the match between this expected thermal
broadening and the observed line width does not rule out the gas also
having some bulk motion, for example from rotation in a disk, it does
show that rotation is not required to explain the data.  Indeed, the
observed line width sets some significant limits on the $v_{\rm orbit}
\sin i$ of the gas.  The expected Keplerian velocity of gas orbiting a
2.2 $M_\sun$ star at 130 AU is 3.8 \kms.  Given the uncertainties in
the range of possible temperatures for the emitting gas, such an
orbital velocity is likely allowed by the data for any $\sin i$.
Orbiting gas at significantly smaller radii, however, would require
values of $\sin i$ significantly less than one.

As noted above, the two measurements of \molecH\ from DoAr 21 by
\citet{Bitner2008} gave different results, suggesting that the bulk of
the 0--0 S(2) emitting gas lies to the north or south of the star.
However, the 5.6 \kms\ FWHM of the S(2) line suggests that this gas is
not coincident with the gas detected by \citet{Bary2002} in the \Sten\
line.  In general the S(2) line traces cooler gas, so the
\citet{Bitner2008} observations may be tracing gas that is somewhat
farther from DoAr 21 than the \citet{Bary2002} observations.

We have presented above a large body of data on DoAr 21, some of it
seemingly contradictory.  To conclude this discussion, we present here
a broad picture of DoAr 21's circumstellar environment.  This is not
the only possible scenario, but we suggest that it is consistent with
all of the available data, and it is testable with future
observations.  In this scenario, DoAr 21 no longer possesses any
substantial orbiting circumstellar material, either in the form of gas
or dust.  Either the system never had much of a disk (possibly as the
result of the formation of the binary companion), or the extreme FUV
and X-ray luminosity have photoevaporated the disk.  The excess
infrared emission, as well as the \molecH\ emission, comes from the
extended material seen in our Gemini images, at hundreds of AU from
the star.  This material may be molecular cloud material that is not
directly associated with the formation of DoAr 21; DoAr 21 may simply
have moved into the vicinity of this gas because of its motion through
the cloud, as we discuss further below.  The FUV flux from DoAr 21 has
created a small-scale PDR in this nearby gas, exciting the observed
\molecH, PAH, and VSG emission.  As in models of PDRs
\citep{Allers2005}, the \Sten\ emission is at the edge of the PDR and
extends farther into the ionized region, explaining why
\citet{Bary2002} detect some emission in an E-W slit that does not
intersect the bright ridge to the north, but \citet{Bitner2008} do not
detect emission in a transition that traces cooler gas.\footnote{While
  the bright spot 1\farcs1 to the N lies outside the slit used by
  \citet{Bary2002}, if we assume a Gaussian profile for the seeing
  with 1\farcs4 FWHM, then roughly 12\% of the flux of a point source
  1\farcs1 away from the star would fall within the slit, so the
  northern ridge could still make some contribution to the observed
  flux.}

One major testable prediction of this scenario is the spatial
distribution of \molecH\ emission.  This could be mapped with
narrow-band filters, or (ideally) with an integral-field spectrograph
that would also measure the line widths and line ratios to probe the
physical conditions in the emitting gas.  For example, observations of
both the S(1) 1--0 and the S(1) 2--1 lines would probe whether the gas
is predominantly thermally excited or pumped by FUV photons.  

Spatially-unresolved observations at high spectral resolution should
detect more flux in a N--S slit than in an E--W one; making observations
close in time with different slit orientations would help determine
whether time variability or spatial distribution is more important.
Given that PDR models predict that \Sten\ emission is spread over a
range of emitting temperatures, the \Sten\ line measured in a N--S slit
may have a stronger flux and a somewhat narrower width than the line
observed by \citet{Bary2002} as it may sample both hotter and cooler
gas.  In contrast, if there is a Keplerian disk that produces the
\Sten\ emission, as suggested by \citet{Bary2002}, then closely-spaced
observations with different slit orientations should produce the same
flux and line width, and mapping of the line emission should show it
closely associated with the stellar position of DoAr 21.

The proper motion of DoAr 21 provides an interesting, though
admittedly speculative, piece of circumstantial evidence in favor of
the hypothesis that DoAr 21 is illuminating ambient cloud material
rather than a circumstellar disk.  \citet{Loinard2008} measure a
proper motion of $(\mu_\alpha \cos \delta, \mu_\delta) = (-26.47 \pm
0.92, -28.23 \pm 0.73)$ mas yr$^{-1}$ for DoAr 21.  This proper motion
in RA is significantly different from the mean proper motion for
$\rho$ Oph members of $(-10 \pm 2, -27 \pm 2)$ mas yr$^{-1}$
\citep{Mamajek2008}, and yet there is ample evidence from the measured
distance, luminosity, Li absorption, and X-ray properties that DoAr 21
is a young star associated with the cloud.  \citet{Mamajek2008} gives
proper motions for 38 candidate $\rho$ Oph members.  Not counting the three
stars that Mamajek rejects as non-members, the only star with an RA
proper motion as discrepant from the mean as DoAr 21's motion is GSS
20 (ROX 7), which \citet{Wilking2005} flag as ``dwarf?'', i.e. a
possible non-member.

If the mean $\rho$ Oph proper motion is representative of the space
motion of the gas near DoAr 21, then DoAr 21 has a tangential motion
of $16.5 \pm 2.2$ mas yr$^{-1}$ ($9.5 \pm 1.3$ km s$^{-1}$) westward
relative to the gas.  This is consistent with the fact that there is
less material visible to the east of DoAr 21 (Fig.\
\ref{figure:gemini-image}) than in other directions.  At this
velocity, DoAr 21 would have been 1\arcsec\ to the east just 60 years
ago, and could have photoevaporated or ionized much of the material in
its vicinity, just as it appears to be doing now in the region within
100 AU of the star.  Such a high velocity relative to the ambient gas
would be unusual, given the typical 1--2 km s$^{-1}$ velocity
dispersions of young associations, but not unheard of.  One example is
the high-velocity young star PV Cep \citep{GoodmanArce2004}, which is
moving at $\sim 20$ \kms\ relative to its local cloud.

To close, we note that the strong PAH emission at large distances from
DoAr 21, and the implied strength of the FUV flux needed to excite it,
lends empirical support to the hypothesis of \citet{Alexander2005}
that there may be significant FUV flux from T Tauri stars even after
accretion has ceased.  Such a flux might enable photoevaporation to
disperse outer disks on a short timescale even after accretion from
the inner disk ceases.

\section{Conclusions}

We have presented a variety of new data on DoAr 21 in an attempt to
understand its circumstellar environment and the influence of
high-energy radiation on it.  High-resolution mid-infrared images show
PAH and very small grain emission in an irregular distribution over
hundreds of AU from the star, with little emission from the inner 100
AU where a circumstellar disk might lie. A new high-resolution X-ray
grating spectrum from \chandra\/ reveals a very luminous, hard corona,
dominated by plasma at temperatures up to $10^8$ K; neither this
spectrum nor new high-resolution optical spectra show any evidence for
accretion.  The inferred FUV radiation from the transition region of
DoAr 21 is strong enough to excite the observed PAH, VSG, and \molecH\
emission, creating a small-scale PDR around DoAr 21.  This strong
radiation may also have played a role in removing any disk from around
DoAr 21, despite its young age.

\acknowledgments

We gratefully acknowledge the support of \chandra\/ grant GO3-4021X,
the National Science Foundation through grant AST-0307830, and
Swarthmore College through a Eugene Lang faculty fellowship.  We thank
the referee for useful comments that improved and clarified the paper.
We are grateful to Victoria Swisher for help with data analysis, and
to Jeff Bary, Tracy Beck, Martin Bitner, Suzan Edwards, Ed Guinan,
Greg Herczeg, Dave Latham, Laurent Loinard, Eric Mamajek, Melissa
McClure, Michael Meyer, Koji Murakawa, Luisa Rebull, Karl Stapelfeldt,
and Russel White for useful discussions.  This research has made use
of the SIMBAD database, operated at CDS, Strasbourg, France, and of
NASA's Astrophysics Data System.


\end{document}